\documentclass[aps,pra,reprint,nofootinbib]{revtex4-2}

\usepackage[utf8]{inputenc}
\usepackage[T1]{fontenc}

\usepackage{aas_macros}

\usepackage{graphicx, epstopdf}

\usepackage{mathtools,amssymb}
\usepackage{tensor, braket}
\usepackage{mathrsfs, stmaryrd, textgreek}
\SetSymbolFont{stmry}{bold}{U}{stmry}{m}{n} 

\usepackage{lmodern,microtype}
\usepackage{booktabs, subcaption}

\usepackage[dvipsnames]{xcolor}
\usepackage[colorlinks=true,allcolors=Blue]{hyperref}
\usepackage[hyphenbreaks]{breakurl}
\usepackage{cleveref}

\let\bs\boldsymbol

\def\core{\text{(1)}}
\def\clad{\text{(2)}}

\let\t\tensor
\let\p\partial
\def\transpose{\mathrm{T}}

\def\dd{\mathrm{d}} 
\let\wop\square 
\def\lc{{\bs \epsilon}} 
\def\conormalN{{N}} 

\def\half{\tfrac{1}{2}}

\def\ihalf{\tfrac{i}{2}}

\def\emA{{A}} 
\def\emF{{F}} 
\def\emG{{G}} 
\def\emP{{\Pi}} 
\def\qmA{{\hat A}} 
\def\qmF{{\hat F}} 
\def\qmG{{\hat G}} 
\def\qmP{{\hat \Pi}} 
\def\permittivity{\text{\textepsilon}} 
\def\permeability{\text{\textmugreek}} 

\def\gopt{{\tilde g}} 
\def\dopt{{\tilde \nabla}} 
\def\Ropt{{\tilde R}} 
\def\wopopt{{\tilde \wop}} 
\def\eopt{{\bs{\tilde \epsilon}}} 

\def\var{\text{\textdelta}} 
\def\formL{{\bs{\mathrm L}}} 
\def\formE{{\bs{\mathrm E}}} 
\def\formP{{\bs{\varpi}}} 
\def\formO{{\bs{\Omega}}} 
\def\formV{{\bs{\varsigma}}} 
\newcommand{\kgbracket}[1]{\braket{\!\braket{#1}\!}} 

\def\phys{\text{(physical)}}
\def\gauge{\text{(gauge)}}
\def\ghost{\text{(ghost)}}
\def\spaceK{\mathfrak K} 
\def\spaceH{\mathfrak H} 

\def\Jj{{\mathscr J}} 
\def\Kk{{\mathscr K}} 

\def\bessel{{\mathfrak b}}
\def\Bessel{{\mathfrak B}}

\def\Wronskian{{\mathfrak W}}

\def\frame{\mathfrak e}
\def\coframe{\mathfrak e}

\def\gravA{{\mathrm g}} 
\def\gravP{{\varphi}} 
\def\aLPI{\alpha\textsubscript{LPI}}

\DeclareMathOperator{\diag}{diag}
\def\cc{\text{c.c.}} 
\def\coherent{\text{coh}} 
\def\ph{\text{ph.}}	
\def\regs{\text{(regular terms)}}
\newcommand{\jump}[1]{\llbracket #1 \rrbracket} 

\usepackage{siunitx}

\begin{document}
\title{Gupta--Bleuler quantization of optical fibers in weak gravitational fields}
\author{Thomas B. Mieling}
\thanks{
ORCID:~\href{https://orcid.org/0000-0002-6905-0183}{0000-0002-6905-0183}\\
E-Mail:~\href{mailto:thomas.mieling@univie.ac.at}{thomas.mieling@univie.ac.at}
}
\affiliation{University of Vienna, Faculty of Physics, Vienna Doctoral School in Physics (VDSP), Vienna Center for Quantum Science and Technology (VCQ) and Research platform TURIS, Boltzmanngasse~5, 1090 Vienna, Austria}

\begin{abstract}
	The theory of gauge-fixed Maxwell equations in linear isotropic dielectrics is developed using a generalization of the standard $R_\xi$ gauge-fixing term. In static space-times, the theory can be quantized using the Gupta--Bleuler method, which is worked out explicitly for optical fibers either in flat space-time or at a constant gravitational potential. This yields a consistent first-principles description of gravitational fiber-optic interferometry at the single-photon level within the framework of quantum field theory in curved space-times.
\end{abstract}

\maketitle

\section{Introduction}

Current experiments concerning gravitational effects on quantum probes utilize massive test particles, exhibiting gravitational phase shifts of the kind first demonstrated by Colella, Overhauser, and Werner \cite{1975PhRvL..34.1472C}.
Experiments of this kind continue to be refined and are now capable of measuring both gravitational accelerations and gravity gradients, see Ref.~\cite{2021QS&T....6b4014T} for an extensive review.
However, as these experiments probe weak gravitational fields with slowly moving massive particles, they are fully explicable in terms of Newtonian gravity \cite{2012CQGra..29v4010Z}. Of course, they can be explained within the framework of general relativity via a weak-field approximation reproducing the predictions of Newtonian gravity, see, e.g., Ref.~\cite{2020PhRvX..10b1014R}, but do not require notions unique to Einstein’s theory of gravitation.

By contrast, the influence of gravity on light is beyond a Newtonian description, as the standard Maxwell equations do not couple to Newton’s gravitational field. Instead, the generally accepted coupling of this kind invokes a generalization of Maxwell’s equations to curved space-times, thereby requiring the framework of general relativity.
While it is possible to express general relativistic corrections to Maxwell’s equations within a Newtonian language by assigning an effective refractive index to the gravitational potential \cite{Schneider:1992}, or by ascribing an effective gravitational mass to the photon \cite{2012CQGra..29v4010Z}, it should be noted that such descriptions are merely reformulations of general relativistic predictions and do not constitute independent theories.

Maxwell’s equations in curved space-times are well-tested at the classical level, e.g., via the Pound--Rebka experiment \cite{1960PhRvL...4..337P}, Shapiro’s time delay experiment \cite{1964PhRvL..13..789S,1968PhRvL..20.1265S}, as well as experiments conducted using the \emph{Gravity Probe A} satellite \cite{1979GReGr..10..181V,1980PhRvL..45.2081V} and eccentric \emph{Galileo} satellites \cite{2018PhRvL.121w1101D}.
Future tests of these equations at the quantum level, in particular using single photons or entangled photon pairs, would constitute the first experimental demonstration of gravitational phase shifts of quantum systems beyond a Newtonian description of gravity \cite{2012CQGra..29v4011R,2017EJPQT.....4....2P,2017NJPh...19c3028H}.

Despite the numerous experimental schemes, which have already been proposed to test gravitational effects on light at the single-photon level \cite{1983PhRvL..51..378T,2012CQGra..29v4011R,2012CQGra..29v4010Z,2017NJPh...19c3028H,2018CQGra..35x4001B,2021arXiv211115591M}, a comprehensive theoretical model of quantum optics in curved space-time geared towards first-principles descriptions of such experiments has been lacking, thus making it necessary to insert the gravitational phase shift \textit{ad hoc}.

A first step in the direction of developing such a theory was done by Anastopoulos and Hu \cite{2021arXiv210612514A}, emphasizing that a consistent first-principles description of quantum optics experiments in gravitational fields necessitates the use of quantum field theory in curved space-times: a framework which was, so far, mainly applied to the regime of strong gravity (Hawking radiation), strong accelerations (Unruh effect), and cosmology.
However, as their model is primarily aimed towards space experiments, it is restricted to light propagation in vacuum, while the aforementioned experimental proposals require the use of optical fibers as a means to implement optical path lengths of the order of hundred kilometers, which are necessary to obtain a measurable gravitational phase shift for laboratory-scale height differences.

The aim of this paper is to fill this gap, by developing a quantum theory of fiber optics in curved space-times, capable of providing a comprehensive theoretical model for the proposed experiments, and assessing future designs.
This theory constitutes a generalization of standard quantum optics in Minkowski space to curved space-times, hence allowing the space-time metric $g$ to differ from the Minkowski metric $\eta$.

From a theoretical perspective, the confinement of photons to a small region by means of optical fibers has the additional advantage of requiring knowledge of the gravitational field in the immediate vicinity of the light rays only. This makes perturbative treatments more transparent, avoiding issues arising in calculations based on infinitely extended plane waves, for which local expansions of the gravitational field are difficult to justify. While this issue is commonly circumvented by using geometric optics, no such approximations are necessary in the setups considered here.

This text is structured as follows.
\Cref{s:general theory} develops the general theory of gauge-fixed Maxwell equations in linear isotropic dielectrics for a class of generalized $R_\xi$ gauge-fixing terms. The analysis of matching conditions at material interfaces shows that only one particular choice of such a gauge-fixing term is compatible with a continuous electromagnetic potential at interfaces.
The quantization of these equations in a static space-time is carried out using the Gupta--Bleuler formalism \cite{1950PPSA...63..681G,Bleuler:1950}.
After this review, \Cref{s:optical fibers} applies these methods to step-index optical fibers (where matching conditions at material interfaces are relevant) both in flat space-time, and in a homogeneous gravitational field.
The results are used to provide a consistent first-principles description of fiber-optic interferometers measuring gravitational phase shifts on single photons and entangled photon pairs, as proposed in the aforementioned references.

Throughout, this document uses geometric units where the gravitational constant $G$, the speed of light in vacuum $c$, the vacuum permeability $\permeability_0$, and the vacuum permittivity $\permittivity_0$ are set to unity. The reduced Planck constant $\hbar$ (with $\sqrt\hbar \approx \SI{1.6163e-35}{m}$) is kept explicit.
Lowercase Latin indices $a$, $b$, $\ldots$ denote abstract indices while Greek letter indices $\mu$, $\nu$, $\ldots$ refer to frame indices.
The signature of the space-time metric tensor $\t g{_a_b}$ is taken to be “mostly positive”, i.e.\ $(-,+,+,+)$.
For two vectors $\t v{^a}$ and $\t w{^a}$, we write $g(v,w) = \t g{_a_b} \t v{^a} \t w{^b}$, while for covectors $\t\phi{_a}$ and $\t\psi{_b}$ we write $g(\phi,\psi) = \t g{^a^b} \t\phi{_a} \t \psi{_b}$, where $\t g{^a^b}$ is the symmetric two-contravariant metric defined by $\t g{^a^c} \t g{_c_b} = \t*\delta{^a_b}.$ Finally, a centered dot indicates contraction of adjacent indices, e.g., $\nabla \cdot v = \t\nabla{_a} \t v{^a}$, $\dd$ denotes the exterior derivative and $\wedge$ the exterior product of differential forms.

\section{General Theory}
\label{s:general theory}

Already at the classical level, there are different ways of formulating Maxwell’s equations, depending on whether the potential $\emA$, or the field strength $\emF$ (or excitation $\emG$) is taken as the fundamental variable.
Similarly, there are multiple approaches to the quantization of electrodynamics, with multiple options available when working with the potential $\emA$, differing in their treatment of the gauge redundancy.

The quantization of the free Maxwell equations in flat space-time in terms of $\emF$ and $\emG$ is described, e.g., in the textbook by Białynicki-Birula \cite{Bialynicki-Birula:1975Q}, and a quantization of optical fiber modes and multi-layered waveguides in flat space-time was developed using this framework by Khrennikov \textit{et al.}~\cite{2012PhyS...85f5404K,2012AIPC.1508..285K}.
However, Dappiaggi and Lang have shown that the quantization based on $\emF$ and $\emG$ can be generalized to curved space-times only under restrictive assumptions on the topology of the space-time manifold \cite{2012LMaPh.101..265D}.
Although such issues are of little concern for the applications of this work, such negative results suggest using different methods applicable to a wider range of problems.

Accordingly, there is a wide class of quantization schemes based on the potential $\emA$ instead.
Such approaches have the further advantage of being particularly well-suited for descriptions of the Aharonov--Bohm effect, see, e.g., Ref.~\cite{DrechslerMayer}, as well as for comparison with models of massive photons, e.g., based on the Proca equation. Similar to the classical theory, the quantization of Maxwell’s equations in terms of the potential requires tools to deal with the gauge redundancy of these equations, for which there are multiple options available.

The reduced-phase-space method commonly used in textbooks on quantum optics \cite{Loudon,GerryKnight,VogelWelsch} proceeds by quantizing the theory in Coulomb gauge, thereby removing all gauge degrees of freedom before quantization.
Although emphasizing the two physical polarization degrees of photons from the outset, this method has the disadvantages of making the gauge-invariance of the theory non-manifest, being based on a preferred reference frame, and rendering the fundamental commutation relations non-trivial due to the presence of constraints, see, e.g., the textbook by Weinberg \cite{Weinberg:1}.

An alternative method was developed by Gupta and Bleuler \cite{1950PPSA...63..681G,Bleuler:1950} (see also the textbook treatments by Itzykson and Zuber \cite{ItzyksonZuber:1980} or Greiner \cite{Greiner:1996}) which is based on an unconstrained quantization of gauge-fixed equations, where superfluous degrees of freedom are removed only after the quantization.
This method has been extended rigorously to curved space-times: Furlani considered static space-times with compact Cauchy surfaces \cite{1995JMP....36.1063F}, and Finster and Strohmaier later allowed for general globally-hyperbolic space-times \cite{2015AnHP...16.1837F}.
While this method requires the use of Krein spaces in intermediate calculations, i.e., function spaces with indefinite pseudo-metrics, such artifacts turn out to be essentially unavoidable if one wishes to maintain covariant equations \cite{Mandrysch:2019}.

While there are further quantization methods available (see, e.g., Folacci \cite{1991JMP....32.2813F}, Sanders--Dappiaggi--Hack \cite{2014CMaPh.328..625S}, or Pfennig \cite{2009CQGra..26m5017P}), this text uses the Gupta--Bleuler quantization method, for the reason of its computational simplicity and manifest covariance, as well as for the aforementioned ease of comparison with the Proca model of massive photons.

\subsection{Maxwell’s Equations}

Maxwell’s equations in Lorentz--Heaviside units are
\begin{align}
	\label{eq:Maxwell general}
	\dd\emF &= 0\,,
	&
	\nabla\cdot\emG + \, j = 0\,,
\end{align}
where $\emF$ is the electromagnetic field strength (Faraday two-form), $\emG$ is the electromagnetic excitation (Maxwell bivector), and $j$ is the current four-vector.
Here, “$\dd$” denotes the exterior derivative, and $\nabla$ is the space-time covariant derivative.

In vacuum, $\emF$ and $\emG$ are related via the space-time metric $\t g{_a_b}$ according to $\t\emG{^a^b} = \t g{^a^c} \t g{^b^d} \t\emF{_c_d}$. Similarly, in a linear isotropic dielectric with four-velocity $\t u{^a}$ (normalized to $\t g{_a_b} \t u{^a} \t u{^b} = -1$), permeability $\permeability$ and permittivity $\permittivity$, the excitation $\emG$ is related to the field strength $\emF$ via
\begin{equation}
	\t \emG{^a^b} = \frac{1}{\permeability} \t\gopt{^a^c} \t\gopt{^b^d} \t\emF{_c_d}\,,
\end{equation}
where $\t\gopt{^a^b}$ is Gordon’s optical metric \cite{1923AnP...377..421G}
\begin{equation}
	\t\gopt{^a^b}
		= \t g{^a^b} + (1-n^2)\t u{^a} \t u{^b}\,,
\end{equation}
with $n = \sqrt{\permeability \permittivity}$ denoting the refractive index.
The covariant optical metric $\t\gopt{_a_b}$, defined as the inverse to $\t\gopt{^a^b}$, is given by
\begin{equation}
	\t\gopt{_a_b}
		= \t g{_a_b} + (1-n^{-2})\t u{_a} \t u{_b}\,,
\end{equation}
where $\t u{_a} = \t g{_a_b} \t u{^b}$. Note that the covariant optical metric $\t\gopt{_a_b}$ is \emph{not} obtained from the covariant optical metric $\t \gopt{^a^b}$ by lowering of indices via the space-time metric $\t g{_a_b}$.
While Gordon’s optical metric has applications in analog gravity, see, e.g., the review by Barceló, Liberati, and Visser \cite{2011LRR....14....3B}, these works typically consider the equations of electromagnetism at the level of the field strength $\t\emF{_a_b}$ and excitation $\t\emG{^a^b}$, but not at the level of the potential $\t\emA{_a}$ as done here.

\subsection{Gauge Fixed Equations}
\label{s:Maxwell equations gauge-fixed}

This section formulates gauge-fixed Maxwell equations in terms of the potential $\t\emA{_a}$ for a family of gauge-fixing terms generalizing the standard $R_\xi$~gauge fixing frequently used in quantum electrodynamics.
The field equations are formulated both using space-time derivatives $\t\nabla{_a}$, i.e., covariant derivatives associated to the space-time metric $\t g{_a_b}$, and optical derivatives $\t\dopt{_a}$ induced by Gordon’s optical metric $\t\gopt{_a_b}$.
Finally, matching conditions for material interfaces are derived.

\subsubsection{Gauge Fixed Lagrangian}
\label{s:gauge-fixed lagrangian}
Maxwell’s equations in a linear isotropic dielectric are variational with the Lagrangian four-form%
\begin{equation}
	\formL_0
		= \left\{
			- \frac{1}{4 \permeability} \t\gopt{^a^b} \t\gopt{^c^d} \t\emF{_a_c} \t\emF{_b_d}
			+ \t j{^a} \t\emA{_a}
		\right\} \lc\,,
\end{equation}
where $\lc$ denotes the space-time volume form.
Gauge fixed equations can be derived from a modified Lagrangian $\formL$, obtained from $\formL_0$ by adding an $R_\xi$~gauge-fixing term
\begin{equation}
	\formL
		= \left\{
			- \frac{1}{4 \permeability} \t\gopt{^a^b} \t\gopt{^c^d} \t\emF{_a_c} \t\emF{_b_d}
			- \frac{1}{2 \xi} \chi_\alpha^2
			+ \t j{^a} \t\emA{_a}
		\right\} \lc\,,
\end{equation}
with the gauge-function
\begin{equation}
	\label{eq:gauge fixing space-time}
	\chi_\alpha
		= n^{1-\alpha}\, \t\nabla{_a}( n^{\alpha - 1} \t\gopt{^a^b} \t\emA{_b})\,,
\end{equation}
for some real parameters $\alpha$ and $\xi$ (which will be chosen later to simplify the equations).
For constant $n$ and covariantly constant $\t u{^a}$ in flat space-time, the gauge conditions $\chi_\alpha = 0$ reduce to the one used by Jauch and Watson in Refs.~\cite{1948PhRv...74..950J,1948PhRv...74.1485J,1949PhRv...75.1249W}.
The gauge condition used by Bei and Liu for the quantization of the electromagnetic field in media of time-dependent refractive indices in flat space-time \cite{2011arXiv1104.2453B} corresponds to $\alpha = 1$. The choice $\alpha = 3$ is frequently used in conjunction with the radiation gauge $\t\emA{_0} = 0$ to obtain a variant of the Coulomb gauge \cite{2010NJPh...12l3008P,SWB-725627441,dorier:tel-02954369}, in which case Gauss’s law arises as a constraint on the space of states \cite{1980NuPhB.163..109R,1980NuPhB.176..477R}. Here, no restriction to a particular value of $\alpha$ is made.

In regions where $\t\gopt{_a_b}$ is sufficiently differentiable to admit a covariant derivative\footnote{This assumption fails at interfaces of materials with different refractive indices, where $\t\gopt{_a_b}$ is discontinuous.} (some later calculations require $\t\gopt{_a_b}$ to be sufficiently differentiable to admit a well-defined curvature tensor), one has
\begin{equation}
	\label{eq:gauge fixing optical}
	\chi_\alpha
		= n^{-\alpha} \t\gopt{^a^b} \t\dopt{_a} (n^\alpha \t\emA{_b})\,.
\end{equation}
Thus, for constant $n$, or for general $n$ but $\alpha = 0$, the condition $\chi_\alpha = 0$ corresponds to a Lorenz gauge formulated using the optical metric (Gordon--Lorenz gauge).
In vacuum, all $\chi_\alpha = 0$ conditions reduce to the standard Lorenz gauge $\t\nabla{^a} \t\emA{_a} = 0$.

\subsubsection{Optical Formulation}

One reason for the specific choice of the above gauge-fixing term is that the field equations become particularly simple when expressed using purely-optical geometric objects, i.e., using the optical metric $\t\gopt{_a_b}$ and the optical covariant derivative $\t\dopt{_a}$.
Using $\t\emF{_a_b} = \t\dopt{_a} \t\emA{_b} - \t\dopt{_b} \t\emA{_a}$, $\lc = n \eopt$ ($\eopt$ denotes the volume-form associated to the optical metric $\t\gopt{_a_b}$), and \cref{eq:gauge fixing optical}, the variation of the Lagrangian form $\formL$ with respect to $\t\emA{_a}$ is
\begin{align}
	\var \formL
		&= \t\formE{^a} \var \t\emA{_a}
		+ \dd(\t\formP{^a} \var \t\emA{_a})\,.
\end{align}
Here, the (vector-valued) Eulerian form $\t\formE{^a}$ and the momentum-form $\t\formP{^a}$ are given by
\begin{align}
	\begin{split}
		\t\formE{^a_b_c_d_e}
		&= \big\{
			\t\dopt{_k} (n \t\emG{^k^a})
			+ n \t j{^a}\\
			&\qquad
			+ \xi^{-1} \t\gopt{^a^k} n^{\alpha} \t\dopt{_k}(n^{1-\alpha} \chi_\alpha)
		\big\} \t\eopt{_b_c_d_e}\,,
	\end{split}
	\\
	\t\formP{^a_b_c_d}
		&= (n \t\emG{^a^e} - n \xi^{-1} \chi_\alpha \t\gopt{^a^e}) \t\eopt{_e_b_c_d}\,.
\end{align}
When expressed solely using optical derivatives, the field equations are thus
\begin{align}
	&\t\dopt{_a} (n \t\emG{^a^b})
			+ \xi^{-1} \t\gopt{^a^b} n^{\alpha} \t\dopt{_a}(n^{1-\alpha} \chi_\alpha)
			+ n \t j{^b}
		= 0\,,
		\\
	\shortintertext{where}
	&\chi_\alpha
		= n^{-\alpha} \t\gopt{^a^b} \t\dopt{_a} (n^\alpha \t\emA{_b})\,.
\end{align}
In the case of a homogeneous material, where both $\permittivity$ and $\permeability$ and hence also $n = \sqrt{\permittivity \permeability}$ are constant, the equations of motion reduce to
\begin{equation}
	\wopopt \t\emA{_a}
	- \t\Ropt{_a^b} \t\emA{_b}
	- (1 - \permeability/\xi) \t\dopt{_a} \chi
	+ \permeability \t\gopt{_a_b} \t j{^b}
	= 0\,,
\end{equation}
where $\chi = \chi_0 \equiv \t\gopt{^a^b} \t\dopt{_a} \t\emA{_b}$, $\wopopt = \t\gopt{^a^b} \t\dopt{_a} \t\dopt{_b}$, and $\t\Ropt{_a_b}$ is the optical Ricci tensor.
In particular, in the Feynman--’t~Hooft gauge $\xi = \permeability$ (which is used in all following calculations), the field equations reduce to
\begin{equation}
	\label{eq:optical wave equation}
	\wopopt \t\emA{_a}
		- \t\Ropt{_a^b} \t\emA{_b}
		+ \permeability \t\gopt{_a_b} \t j{^b}
		= 0\,.
\end{equation}

\subsubsection{Space-Time Formulation}

The optical formulation is well-suited for the description of the electromagnetic field in homogeneous isotropic linear media, see for example the derivation of the Frank--Tamm formula for Cherenkov radiation by Jauch and Watson \cite*{1948PhRv...74.1485J}.
However, this formulation is inadequate at interfaces of such media, as discontinuities of the optical metric render the optical derivative $\t\dopt{_a}$ and the optical curvature $\t\Ropt{_a^b}$ ill-defined there.
It thus behooves to analyze interface conditions without appealing to these objects.
In terms of the space-time derivative $\t\nabla{_a}$, the variation of the Lagrangian with respect to $\t\emA{_a}$ reads
\begin{align}
	\label{eq:form L s-t}
	\var \formL
		&= \t\formE{^a} \var\t\emA{_a} + \dd(\t\formP{^a} \var\t\emA{_a})\,,
		\\
	\shortintertext{where}
	\label{eq:form E s-t}
	\begin{split}
		\t\formE{^a_b_c_d_e}
		&= \big\{
			\t\nabla{_k} \t\emG{^k^a}
			+ \t j{^a}\\
			&\qquad
			+ \xi^{-1} \t\gopt{^a^k} n^{\alpha - 1} \t\nabla{_k}(n^{1 - \alpha} \chi_\alpha)
		\big\} \t\lc{_b_c_d_e}\,,	
	\end{split}
	\\
	\label{eq:form P s-t}
	\t\formP{^a_b_c_d}
		&= (\t\emG{^a^e} - \xi^{-1} \chi_\alpha \t\gopt{^a^e}) \t\lc{_e_b_c_d}\,.
\end{align}
Expressed purely using space-time derivatives, the field equations are thus
\begin{align}
	\label{eq:EOM gauge-fixed general}
	&\t\nabla{_a} \t\emG{^a^b}
		+ \xi^{-1} \t\gopt{^a^b} n^{\alpha -1} \t\nabla{_a}(n^{1-\alpha} \chi_\alpha)
		+ \t j{^b}
	= 0\,,
	\\
	\shortintertext{where}
	&\chi_\alpha
		= n^{1-\alpha}\, \t\nabla{_a}( n^{\alpha - 1} \t\gopt{^a^b} \t\emA{_b})\,.
\end{align}
These equations do not rely on the optical derivative $\t\dopt{_a}$ and are thus meaningful also at material interfaces where the optical metric $\t\gopt{_a_b}$ is discontinuous.

\subsubsection{Evolution of the Gauge Function}
\label{s:gauge function evolution}

Taking the divergence of \cref{eq:EOM gauge-fixed general}, using $\t\nabla{_a} \t\nabla{_b} \t\emG{^a^b} = 0$ as well as the continuity equation $\t\nabla{_a} \t j{^a} = 0$, yields
\begin{equation}
	\label{eq:gauge wave equation general}
	\t\nabla{_a}(n^{\alpha -1} \t\gopt{^a^b} \t\nabla{_b}(n^{1- \alpha} \chi_\alpha)) = 0\,,
\end{equation}
or when formulated in terms of the optical derivative:
\begin{equation}
	\label{eq:gauge wave equation optical}
	\t\dopt{_a}(n^{\alpha} \t\gopt{^a^b} \t\dopt{_b}(n^{1- \alpha} \chi_\alpha)) = 0\,.
\end{equation}
In regions of constant $n$, all gauge functions $\chi_\alpha$ satisfy the optical wave equation
\begin{equation}
	\label{eq:gauge wave equation optical constant n}
	\wopopt \chi_\alpha
		\equiv \t\gopt{^a^b}\t\dopt{_a}\t\dopt{_b} \chi_\alpha
		= 0
		\qquad\text{(for constant $n$)}\,.
\end{equation}
For arbitrary $n$, one has the important special cases $\alpha = 0$ and $\alpha = 1$ (cf.\ \Cref{s:gauge-fixed lagrangian})
\begin{align}
	\label{eq:gauge wave equation special}
	\t\gopt{^a^b}\t\dopt{_a}\t\dopt{_b} (n \chi_0) &= 0\,,
	&
	\t\nabla{_a}(\t\gopt{^a^b} \t\nabla{_b} \chi_1) &= 0\,.
\end{align}
The equations \eqref{eq:gauge wave equation general} and \eqref{eq:gauge wave equation optical} come from the variation of the Lagrangian four-form
\begin{equation}
	\formL_\alpha'
		= - \half n^{\alpha-1} \t\gopt{^a^b} \t\nabla{_a}(n^{1 - \alpha} \chi_\alpha) \t\nabla{_b}(n^{1 - \alpha} \chi_\alpha) \lc \,,
\end{equation}
with associated Eulerian four-form $\formE'_\alpha$ and momentum three-form $\formP'_\alpha$ given by
\begin{align}
	\t{{\formE'}}{_\alpha_c_d_e_f}
		&= n^{1 - \alpha}\t\nabla{_a} [n^{\alpha-1} \t\gopt{^a^b}\t\nabla{_b}(n^{1 - \alpha} \chi_\alpha)]
		\t\lc{_c_d_e_f}\,,
		\\
	\t{{\formP'}}{_\alpha_c_d_e}
		&= - \t\gopt{^a^b} \t\nabla{_a}(n^{1-\alpha} \chi_\alpha) \t\lc{_b_c_d_e}\,.
\end{align}
The special case $\alpha = 1$ makes these expressions particularly simple, while $\alpha = 3$ ensures that the refractive index enters the equations only with the powers $n^2 = \permittivity\permeability$ and $n^{-2} = 1/(\permittivity\permeability)$, i.e., without square roots of either $\permeability$ or $\permittivity$.

\subsection{Klein--Gordon Product}
\label{s:KG product}

For two solutions $\t\emA{_a}$ and $\t{{\emA'}}{_a}$ with associated momentum-forms $\t\formP{^a}$ and $\t{{\formP'}}{^a}$, define the three-form%
\begin{equation}
	\formO(A, A')
		= \t*\emA{^*_a} \t{{\formP'}}{^a}
		- \t{{\emA'}}{_a} \t\formP{^*^a}\,,
\end{equation}
where a star denotes complex conjugation.
By definition of $\formP'$, one has $\dd(A \cdot \formP') = \var_A \formL(A') - A \cdot \formE(A')$, where the second term vanishes if $A'$ satisfies the equations of motion. Consequently, one has the “on-shell” formula
\begin{equation}
	\dd \formO(A, A')
		= \var_\emA \formL(\emA') - \var_{\emA'} \formL(\emA)\,.
\end{equation}
In the absence of sources, the Lagrangian $\formL$ is quadratic in $A$, so $\var_A \formL(A')$ is symmetric under interchange of $A$ and $A'$, hence $\formO$ is closed.
It follows (under the assumption of global hyperbolicity and suitable decay of the fields at large distances) that the integral of $\formO$ over any Cauchy surface $\Sigma$ is independent of the choice of such surface.
Hence, the Klein--Gordon product $\kgbracket{A | A'}$, defined as
\begin{equation}
	\kgbracket{A | A'}
		= i \int_\Sigma \formO(A, A')\,,
\end{equation}
is independent of the choice of Cauchy surface $\Sigma$ used to evaluate the integral.
Given any such surface, denote by $\conormalN$ its future-pointing unit conormal, i.e., the unit normal one-form with the sign chosen such that $\conormalN(X) < 0$ for every future-pointing timelike vector field $X$, and by $\formV$ the induced volume form. On $\Sigma$, the space-time volume form factorizes as $\lc = - \conormalN \wedge \formV$, and the vector-valued three-form $\formP$ factorizes as $\formP = \emP \otimes \formV$, where $\emP$ is the vector field
\begin{equation}
	\label{eq:GB momentum}
	\t \emP{^a} = - (\t\emG{^a^b} - \xi^{-1} \chi_\alpha \t\gopt{^a^b}) \t\conormalN{_b}\,,
\end{equation}
which generalizes the expression used by Jauch and Watson for homogeneous dielectrics in flat space-time \cite{1948PhRv...74.1485J}.
With this notation, the Klein--Gordon product takes the form
\begin{equation}
	\label{eq:KG product EM general}
	\kgbracket{A | A'}
		= i \int_\Sigma (\t\emA{_a}^* \t{{\emP'}}{^a} - \t{{A'}}{_a} \t\emP{^a}^*) \formV\,.
\end{equation}

By the same method, one can construct a Klein--Gordon product for the gauge functions $\chi_\alpha$, which yields
\begin{equation}
	\label{eq:KG product gauge general}
	\kgbracket{\chi | \chi'}_\alpha
		= i \int_\Sigma [\chi^* \t\nabla{_a} (n^{1-\alpha}\chi') - \chi' \t\nabla{_a}(n^{1 - \alpha} \chi^*)] \t\gopt{^a^b} \t\conormalN{_a} \formV\,.
\end{equation}
In vacuum, \cref{eq:KG product gauge general} reduces to the standard Klein--Gordon product for a free scalar field \cite{Wald:1994}, and if, moreover, the gauge condition is satisfied, $\chi = 0$, the product given in \cref{eq:KG product EM general} reproduces the standard “symplectic product” of the free electromagnetic field \cite{Wald:1994,Fursaev:2011}.

\subsection{Interface Conditions}
\label{s:inteface conditions}

This section describes how solutions obtained in two regions of space-time must match at interfaces of materials with different refractive indices to form one global solution satisfying the field equations everywhere (in the distributional sense).

\subsubsection{Maxwell’s Equations}

At material interfaces, Maxwell’s equations imply continuity conditions for some components of the electromagnetic field, and relate discontinuities of other components to surface charges and surface currents. Specifically, denoting the jump of a quantity $x$ at the interface by $\jump{x}$, Maxwell’s equations imply $\jump{F \wedge \nu} = 0$ and $\jump{\emG \cdot \nu} = \jmath$, where $\nu$ is the unit conormal to the interface, and $\jmath$ is the four-vector of surface charges and surface currents.

To demonstrate these equations, it is advantageous to adopt local coordinates $(z, \t\zeta{^\mu})$, where $\mu$ ranges from zero to three, such that the interface is given by $z = 0$, and the conormal is $\nu = \dd z$. In these coordinates, $\dd F = 0$ implies
\begin{equation}
	\t\p{_z} \t\emF{_\mu_\nu}
	+ \t\p{_\mu} \t\emF{_\nu_z}
	+ \t\p{_\nu} \t\emF{_z_\mu} = 0\,.
\end{equation}
If the tangential derivatives are regular functions, i.e., containing neither Dirac-delta terms located at $z = 0$ nor derivatives thereof, then $\t\emF{_\mu_\nu}$ must be continuous (for otherwise the left-hand side would contain an unbalanced Dirac-delta term). In coordinate-independent notation, continuity of the tangential components $\t\emF{_\mu_\nu}$ is expressed as continuity of $\emF \wedge \nu$.

The second condition follows from the equation $\nabla \cdot \emG + j = 0$. In the adapted coordinates described above, the surface current $j$ (which is assumed to be tangential) has the form ${\t j{^\mu} = \t\jmath{^\mu} \delta(z)}$, where $\t\jmath{^\mu}$ is a regular vector field defined on the interface. Assuming, as above, $\emG$ and its transverse derivatives to be regular, the inhomogeneous field equations yield
\begin{equation}
	\t\p{_z} \t\emG{^z^\mu} + \regs + \t\jmath{^\mu} \delta(z) = 0\,,
\end{equation}
and so $\t\emG{^\mu^z}$ must jump by $\t\jmath{^\mu}$, i.e., $\emG \cdot \nu$ jumps by $\jmath$.

In case one works with a potential $\emA$, a further condition is obtained from the requirement that $\emF = \dd A$ is regular. Specifically, one has
\begin{equation}
	\t\emF{_z_\mu} = \t\p{_z}\t\emA{_\mu} - \t\p{_\mu}\t\emA{_z}\,,
\end{equation}
and if the transverse derivatives of $\t\emA{_z}$ are regular (they may be discontinuous), then the condition that $\t\emF{_z_\mu}$ be regular requires $\t\emA{_\mu}$ to be continuous, so $\emA \wedge \nu$ must be continuous.
This completes the derivation of the interface conditions listed in \Cref{tab:matching Maxwell}, based on Maxwell’s equations \emph{without} gauge-fixing terms.

\begin{table}[t]
	\centering
	\begin{tabular}{@{}lr@{}}
		\toprule
		Condition	& Jump condition\\
		\midrule
		Regular field strength
			& $\jump{\emA \wedge \nu} = 0$
		\\
		Homogeneous field equation
		& $\jump{\emF \wedge \nu} = 0$
		\\
		Inhomogeneous field equation
		& $\jump{\emG \cdot \nu} = \jmath$
		\\
		\bottomrule
	\end{tabular}
	\caption{Matching conditions for Maxwell’s equations at an interface with unit conormal $\nu$. Here, $\jump{x}$ denotes the jump of a quantity $x$, and $\jmath$ is the tangential four-vector of surface charges and surface currents.}
	\label{tab:matching Maxwell}
\end{table}

\subsubsection{Gauge-Fixed Equations}

\begin{table}[b]
	\centering
	\begin{tabular}{@{}lr@{}}
		\toprule
		Condition	& Jump condition\\
		\midrule
		Regular gauge function
			& $\jump{ n^{\alpha - 1} \gopt(A, \nu) } = 0$
		\\
		Regular field strength
			& $\jump{\emA \wedge \nu} = 0$
		\\
		Homogeneous field equation
			& $\jump{\emF \wedge \nu} = 0$
		\\
		Transverse inhomogeneous field equation
			& $\jump{\emG \cdot \nu} = \jmath$
		\\
		Normal inhomogeneous field equation
			& $\jump{ n^{1 - \alpha} \chi_\alpha } = 0$
		\\
		\bottomrule
	\end{tabular}
	\caption{Matching conditions for the gauge-fixed equations. Notation as in \Cref{tab:matching Maxwell}.}
	\label{tab:matching gauge-fixed general}
\end{table}

For the gauge-fixed field equations \eqref{eq:EOM gauge-fixed general}, the matching condition differ from those of the standard Maxwell equations.
The $z$-component of \cref{eq:EOM gauge-fixed general} yields
\begin{equation}
	\label{eq:interface gauge-fixed derivation1}
	\xi^{-1} \t\gopt{^z^z} n^{\alpha - 1} \t\p{_z}(n^{1-\alpha} \chi_\alpha) + \regs = 0\,.
\end{equation}
From this, one obtains two conditions.
First, $\chi_\alpha$ must be regular, for otherwise its $z$-derivative would produce an unbalanced derivative of a Dirac-delta distribution (and, moreover, the product $n^{1-\alpha} \chi$ would be ill-defined except for $\alpha = 1$). Since $\chi_\alpha = n^{1-\alpha}\t\p{_z}( n^{\alpha - 1} \t\gopt{^z^a} \t\emA{_a})$, this entails that $n^{\alpha - 1} \t\gopt{^z^a} \t\emA{_a}$ must be continuous, so $\jump{n^{\alpha - 1} \gopt(\emA, \nu)} = 0$.
Second, \cref{eq:interface gauge-fixed derivation1} implies that $n^{1-\alpha} \chi_\alpha$ is continuous, for otherwise the $z$-derivative would produce a Dirac-delta term which cannot be canceled by the remaining regular terms.

Having established that the second term in \cref{eq:EOM gauge-fixed general} is regular, the remaining analysis of interface conditions implied by this equation proceeds exactly as for the original Maxwell equations without the gauge-fixing term.
The resulting interface conditions are summarized in \Cref{tab:matching gauge-fixed general}.

\subsubsection{Continuous Potentials}

Of special importance is the choice $\alpha = 1$ in the gauge function $\chi$, defined in \cref{eq:gauge fixing space-time,eq:gauge fixing optical}.
In this case, $\gopt(\emA, \nu)$ is continuous. If, moreover, $\nu$ is orthogonal to the four-velocity $u$ of the medium, then $\gopt(\cdot, \nu) = g(\cdot, \nu)$, so $g(\emA,\nu)$ is continuous. Combined with the continuity of the tangential components $\t\emA{_\mu}$, one thus finds that \emph{all} components of $\emA$ are continuous in this case.
Under the condition $\t u{^a} \t\nu{_a} = 0$, the choice $\alpha = 1$ thus leads to the matching conditions listed in \Cref{tab:matching gauge-fixed special}.

\begin{table}[t]
	\centering
	\begin{tabular}{@{}lr@{}}
		\toprule
		Condition	& Jump condition\\
		\midrule
		Continuous potential
			& $\jump{\emA} = 0$
		\\
		Homogeneous field equation
			& $\jump{\emF \wedge \nu} = 0$
		\\
		Inhomogeneous field equation
			& $\jump{\emG \cdot \nu} = \jmath$
		\\
		Continuous gauge function
			& $\jump{ \chi_1 } = 0$
		\\
		\bottomrule
	\end{tabular}
	\caption{Matching conditions for $\alpha = 1$ and $\t u{^a} \t\nu{_a} = 0$, where $\t u{^a}$ is the four-velocity of the medium and $\t \nu{_a}$ is the unit conormal to the interface. $\jump{x}$ denotes the jump of a quantity $x$ at the interface.}
	\label{tab:matching gauge-fixed special}
\end{table}

\subsection{Gupta--Bleuler Quantization}

This section contains a general outline of the Gupta--Bleuler quantization method, which is carried out in more detail for concrete applications in \Cref{s:optical fibers}.

\subsubsection{The Classical Picture}

Before considering the quantized theory, it is useful to consider the classical theory to develop some nomenclature.

Due to the presence of gauge symmetry, Maxwell’s equations \eqref{eq:Maxwell general} do not provide a well-posed Cauchy problem for the potential $\emA$: if $\emA'$ is any solution for Maxwell’s equations with prescribed initial data $(\emA|_\Sigma, \dot\emA|_\Sigma)$, then another solution is given by $\emA'' = \emA' + \dd \lambda$, where $\lambda$ is any smooth function supported away from the initial Cauchy surface $\Sigma$.
By contrast, the gauge-fixed equations described in \Cref{s:Maxwell equations gauge-fixed} \emph{do} provide a well-posed Cauchy problem.
Considering these equations at face value, i.e., \emph{without} imposing the condition $\chi = 0$ from the outset, the gauge-fixed equations are free of constraints, so that initial data can be prescribed arbitrarily.
However, not every solution the gauge-fixed equations obtained in this fashion will also solve the original Maxwell equations.

\begin{figure}[b]
	\centering
	\includegraphics[width=\columnwidth]{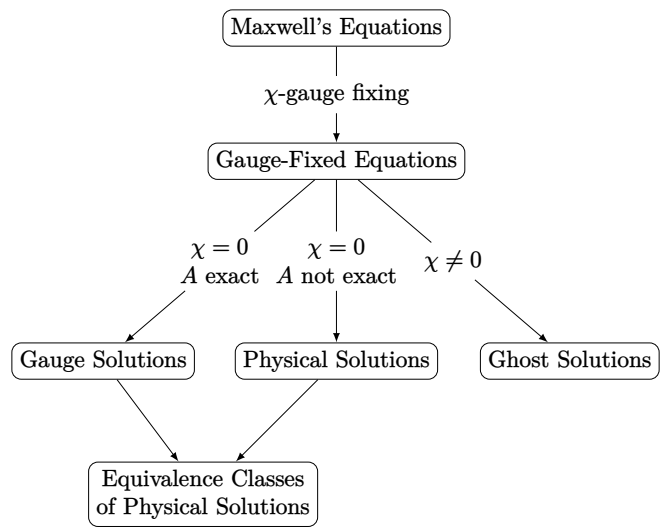}
	\caption{Solution scheme for Maxwell’s equations in terms of the potential $\emA$.}
	\label{fig:flow diagram classical}
\end{figure}

To describe this situation, it is useful to introduce the following nomenclature for solutions to the gauge-fixed equations, adapted to standard terminology of the quantized theory:
\begin{itemize}
	\item A solution with $\chi \neq 0$ is referred to as a \emph{ghost solution} (sometimes called \emph{unphysical}) as it does not satisfy Maxwell’s equations.
	\item A solution with $\chi = 0$ but with $\emA$ being exact, i.e., the gradient of a function, is said to be a \emph{gauge solution}, as it satisfies Maxwell's equations simply because $\emF$ and $\emG$ both vanish (as do all holonomy integrals). Such solutions describe trivial connections.
	\item Finally, solutions with $\chi = 0$ and $\emA$ being not exact are referred to as \emph{physical solutions}, as they are non-trivial solutions to the original Maxwell equations.
\end{itemize}
Since all physical solutions within the same cohomology class are physically indistinguishable (as the fields $\emF$, $\emG$, and all holonomy integrals agree), the set of physically distinguishable solutions is given by the cohomology classes of physical solutions, i.e., “physical solutions up to gauge solutions.” This scheme is visualized in \Cref{fig:flow diagram classical}.

It is worth noting that if the gauge function $\chi$ satisfies a hyperbolic equation (cf.\ \Cref{s:gauge function evolution}, where difficulties may arise from interfaces where the optical metric is discontinuous), then the distinction between $\chi = 0$ and $\chi \neq 0$ can be made at the level of initial data prescribed on any Cauchy surface, as $\chi$ vanishes globally if and only if its initial data is trivial.

\subsubsection{The Quantum Picture}

Similar to the fact that the gauge-fixed Maxwell equations, \emph{without} the gauge condition $\chi = 0$ enforced, are free of constraints and thus allow for arbitrary initial data, the first step of the Gupta--Bleuler quantization method is an unconstrained quantization of the gauge-fixed equations.
In a static space-time, this yields quantum field operators $\qmA$ and $\qmP$ satisfying the Heisenberg commutation relation
\begin{equation}
	[\t\qmA{_a}(t, \bs x), \t\qmP{^b}(t, \bs y)]
		= i\hbar \t*\delta{_a^b} \delta(\bs x - \bs y)\,,
\end{equation}
acting on a space of quantum states $\spaceK$. This space $\spaceK$ contains a vacuum state $\ket 0$, which is annihilated by all lowering-operators $\hat a(\psi)$ corresponding to classical solutions $\psi$ of positive frequency.
However, $\spaceK$ is not a Hilbert space but a Krein space, carrying an indefinite inner product, related to the indefiniteness of the Klein--Gordon product of solutions to the gauge-fixed equations, defined in \cref{eq:KG product EM general}.

\begin{figure}[t]
	\centering
	\includegraphics[width=\columnwidth]{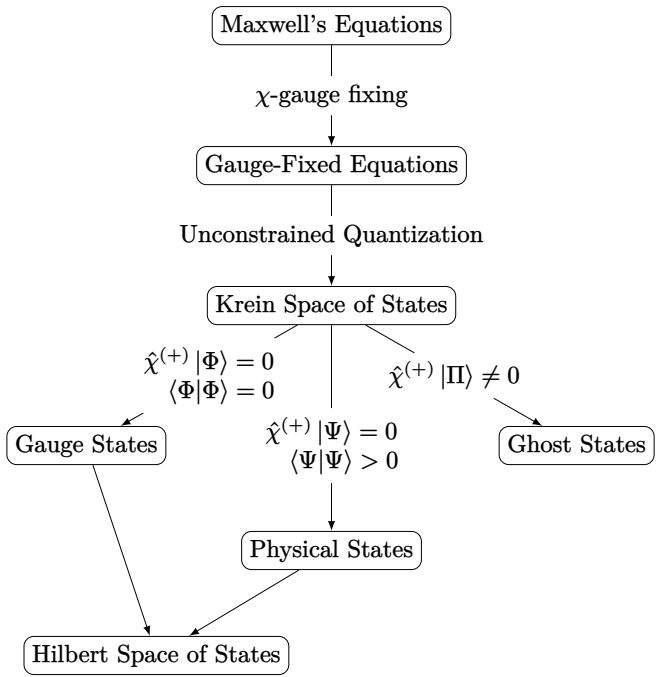}
	\caption{Schematic overview of the Gupta--Bleuler quantization method.}
	\label{fig:flow diagram GB}
\end{figure}

The indefiniteness of the inner product on $\spaceK$ is deeply tied to the presence of ghost and gauge solutions. For example, classical gauge solutions have $\chi_\alpha = 0$ and $\emG = 0$, hence $\emP = 0$, and thus a vanishing Klein--Gordon norm. This gives rise to gauge states of vanishing pseudo-norm in $\spaceK$.

The distinction of physical and unphysical classical modes by means of the gauge function $\chi$ is carried over to the quantum theory by regarding as physical only those states $\ket\Psi$, which satisfy the Gupta--Bleuler condition
\begin{equation}
	\hat \chi^{(+)} \ket\Psi = 0\,.
\end{equation}
Here, $\hat \chi^{(+)}$ is the quantum operator associated to the positive-frequency part of the gauge-function $\chi$, i.e., the part whose temporal Fourier transform contains contributions of the form $e^{-i \omega t}$ with $\omega > 0$ only.

Similar to the categorization of classical solutions introduced above, one uses the following classification of quantum states in $\spaceK$:
\begin{itemize}
	\item \emph{Ghost} states $\ket \Pi$ are those violating the Gupta--Bleuler condition.
	\item \emph{Gauge} states $\ket\Phi$ satisfy the Gupta--Bleuler condition but have \emph{vanishing} pseudo-norm.
	\item \emph{Physical} states $\ket\Psi$ are those satisfying the Gupta--Bleuler condition and having \emph{positive} pseudo-norm.
\end{itemize}

The Gupta--Bleuler quantization scheme requires that there are no states other than these three categories, i.e., no states satisfying the Gupta--Bleuler condition but having negative norm.
Numerical tests indicate that no such states arise in the applications considered here.

Similar to the classical picture, the space of physically distinguishable states is then obtained by taking the quotient of physical states up to gauge states, which yields a Hilbert space $\spaceH$ carrying a positive-definite inner product.
This procedure is visualized in \Cref{fig:flow diagram GB}.

One may inquire the necessity of introducing the seemingly superfluous gauge and ghost modes in the theory, if they are to be discarded later in the quantization process. However, one finds that such modes are \emph{necessary} to obtain covariant gauge formulations of the quantized electromagnetic field even in vacuous Minkowski space-time \cite{Mandrysch:2019}.

\section{Optical fibers}
\label{s:optical fibers}

In this section, the Gupta--Bleuler quantization scheme is carried out for step-index optical fibers. The problem is considered first in flat space-time, and the solution is generalized later to optical fibers held at a constant potential in a weak gravitational field.

Contrary to infinitely extended homogeneous media or vacuum in flat space-time, the gauge and ghost modes turn out to have dispersion relations different from the physical modes.

\subsection{Optical Fibers in Flat Space-Time}
\label{s:fibers in flat space}

Consider a cylindrical step-index optical fiber which has constant refractive index $n_1$ in its core ($r < \rho$) and constant refractive index $n_2$ in its cladding ($r > \rho$) with $n_1 > n_2$.
In what follows, cylindrical coordinates $r, \theta, z$ will be used, where $z$ measures distance along the symmetry axis of the optical fiber at rest, which thus has four-velocity $\t\p{_t}$.
It is convenient to use the complex frame $\t\frame{_\mu}$ and coframe $\t\coframe{^\mu}$ defined as
\begin{align}
	\label{eq:fiber complex frame}
	\t\frame{_t}
		&= \t\p{_t}\,,
	&
	\t\frame{_\parallel}
		&= \t\p{_z}\,,
	&
	\t\frame{_\pm}
		&= \tfrac{1}{\sqrt 2}(\t\p{_r} \pm \tfrac{i}{r} \t\p{_\theta})\,,
	\\
	\label{eq:fiber complex coframe}
	\t\coframe{^t}
		&= \dd t\,,
	&
	\t\coframe{^\parallel}
		&= \dd z\,,
	&
	\t\coframe{^\pm}
		&= \tfrac{1}{\sqrt 2}( \dd r \mp i r \dd \theta)\,.
\end{align}
In the following, all calculation will be carried out in the Feynman--’t~Hooft gauge\footnote{As the gauge-fixing parameter $\xi$ was assumed to be constant, the use of the Feynman--’t~Hooft gauge requires the considered material to have a constant permeability $\permeability$. This is satisfied by typical optical fibers, as they are non-magnetic and thus have $\permeability = 1$.} for the $R_{\xi,1}$-gauge fixing function:
\begin{align}
	\xi &= \permeability\,,
	&
	\alpha &= 1\,,
\end{align}
see \Cref{s:gauge-fixed lagrangian} for definitions.

\subsubsection{Gauge Potential}

Separating Fourier modes of different frequency $\omega$, propagation constant $\beta$, and azimuthal mode index $m$, one is led to the ansatz
\begin{equation}
	\label{eq:potential decomposition}
	\emA = (
			\t a{_t} \t\coframe{^t}
			+ \t a{_+} \t\coframe{^+}
			+ \t a{_-} \t\coframe{^-}
			+ \t a{_\parallel} \t\coframe{^\parallel}
		) e^{i(\beta z + m \theta - \omega t)}\,,
\end{equation}
where the functions $\t a{_t}$, $\t a{_\parallel}$, $\t a{_\pm}$ depend on the radial coordinate $r$ only.
The gauge-fixed equations~\eqref{eq:optical wave equation} without external sources, valid in the core and cladding separately, are
\begin{subequations}
\label{eq:fiber radial equations}
\begin{align}
	\t\Bessel{_m} \t a{_t} &= 0\,,
	&
	\t\Bessel{_{m+1}} \t a{_+} &= 0\,,
	\\
	\t\Bessel{_m} \t a{_\parallel} &= 0\,,
	&
	\t\Bessel{_{m-1}} \t a{_-} &= 0\,,
	&
\end{align}
\end{subequations}
where $\t\Bessel{_\nu}$ is the Bessel operator
\begin{equation}
	\t\Bessel{_\nu} = r^2 \t*\p{_r^2} + r \t\p{_r} + r^2 (n^2 \omega^2 - \beta^2) - \nu^2 \,.
\end{equation}
Requiring the field to be regular on the symmetry axis and to decay at large radii, the solutions, valid in the core and cladding separately, are found to be
\begin{subequations}
\begin{align}
	\t a{_t}(r) &= f_m(q^\core_t, q^\clad_t, r)\,, \\
	\t a{_\parallel}(r) &= f_m(q^\core_\parallel, q^\clad_\parallel, r)\,, \\
	\t a{_+}(r) &= f_{m+1}(q^\core_+, q^\clad_+, r)\,, \\
	\t a{_-}(r) &= f_{m-1}(q^\core_-, q^\clad_-, r)\,,
\end{align}
\end{subequations}
where the coefficients $q_\mu^\core$ and $q_\mu^\clad$ are constants, and the functions $f_\nu(q, q', r)$ are given in terms of Bessel functions of the first kind, $J_\nu$, and modified Bessel functions of the second kind, $K_\nu$, as
\begin{equation}
	f_\nu(q, q', r) = \begin{cases}
		q\phantom{'} J_\nu(U r/\rho)	&	r < \rho\,,
		\\
		q' K_\nu(W r/\rho)	&	r > \rho\,.
	\end{cases}
\end{equation}
Here, $U$ and $W$ are defined as
\begin{subequations}
\label{eq:fibers U W}
\begin{align}
	U &= \sqrt{+\rho^2(n_1^2 \omega^2 - \beta^2)}\,,
	\\
	W &= \sqrt{-\rho^2(n_2^2 \omega^2 - \beta^2)}\,,
\end{align}
\end{subequations}
where it is assumed that $n_1^2 \omega^2 < \beta^2 < n_2^2 \omega^2$, for otherwise there would be no guided modes.
The gauge function $\chi = \t \gopt{^a^b} \t\dopt{_a} \t\emA{_b}$ takes the form
\begin{equation}
	\label{eq:fiber gauge function}
	\chi
		= f_m(q_\chi, q'_\chi,r) e^{i(\beta z + m \theta - \omega t)}\,,
\end{equation}
where
\begin{align}
	q_\chi
		&= i (n_1^2 \omega q_t^\core + \beta q_\parallel^\core) + \tfrac{U}{\sqrt{2}\rho}(q_+^\core - q_-^\core)\,,
	\\
	q'_\chi
		&= i (n_2^2 \omega q_t^\clad + \beta q_\parallel^\clad) - \tfrac{W}{\sqrt{2}\rho}(q_+^\core + q_-^\clad)\,.
\end{align}

\subsubsection{Canonical Momentum}
The general formula for the momentum density in \cref{eq:GB momentum} yields
\begin{align}
	\label{eq:fiber momentum}
	\t \emP{^a}
		&= \t\emG{^a^0} - \chi \t\gopt{^a^0}\,.
\end{align}
Contrary to the components of the potential $\emA$, $\t\Pi{^+}$ contains Bessel functions of order $m - 1$, while $\t \Pi{^-}$ contains those of order $m + 1$. This is because the metric-equivalent one-form to the vector $\t\frame{_\pm}$ is $\t\coframe{^\mp}$. For this reason it is convenient to work with the complex-conjugate field
\begin{equation}
	\label{eq:momentum decomposition}
	\emP^* = (
		\t{\bar\pi}{^t} \t\frame{_t}
		+ \t{\bar\pi}{^\parallel} \t\frame{_\parallel}
		+ \t{\bar\pi}{^+} \t\frame{_+}
		+ \t{\bar\pi}{^-} \t\frame{_-}
	)e^{-i(\beta z + m \theta - \omega t)} \,.
\end{equation}
Here, the functions $\t{\bar\pi}{^\mu}$ take the form
\begin{subequations}
\begin{align}
	\t{\bar\pi}{^t}(r)
		&= f_{m} (p^\core_t, p^\clad_t, r )\,,
		&
	\t{\bar\pi}{^+}(r)
		&= f_{m+1} (p^\core_+, p^\clad_+, r )\,,
		\\
	\t{\bar\pi}{^\parallel}(r)
		&= f_{m} (p^\core_\parallel, p^\clad_\parallel, r )\,,
		&
	\t{\bar\pi}{^-}(r)
		&= f_{m-1} (p^\core_-, p^\clad_-, r )\,,
\end{align}
\end{subequations}
where the coefficients $p^\core_\mu$, $p^\clad_\mu$ are expressible in terms of the coefficients $q^\core_\mu$, $q^\clad_\mu$ as
\begin{subequations}
\begin{align}
	p^\core_t
		&= - i n_1^2 [n_1^2 \omega \bar q^\core_t + \beta \bar q^\core_\parallel] + \tfrac{n_1^2 U}{\sqrt 2 \rho} [\bar q^\core_+ - \bar q^\core_-] \,,
		\\
	p^\core_\parallel
		&= + i n_1^2 [\beta \bar q^\core_t + \omega \bar q^\core_\parallel] \,,
		\\
	p^\core_+
		&= + n_1^2 [\tfrac{U}{\sqrt 2 \rho} \bar q^\core_t + i \omega \bar q^\core_+] \,,
		\\
	p^\core_-
		&= - n_1^2 [\tfrac{U}{\sqrt 2 \rho} \bar q^\core_t - i \omega \bar q^\core_-] \,,
		\\
	p^\clad_t
		&= -i n_2^2 [n_2^2 \omega \bar q^\clad_t + \beta \bar q^\clad_\parallel] - \tfrac{n_2^2 W}{\sqrt 2 \rho} [\bar q^\clad_+ + \bar q^\clad_-] \,,
		\\
	p^\clad_\parallel
		&= + i n_2^2 [\beta \bar q^\clad_t + \omega \bar q^\clad_\parallel] \,,
		\\
	p^\clad_+
		&= + n_2^2 [\tfrac{W}{\sqrt 2 \rho} \bar q^\clad_t + i \omega \bar q^\clad_+] \,,
		\\
	p^\clad_-
		&= + n_2^2 [\tfrac{W}{\sqrt 2 \rho} \bar q^\clad_t + i \omega \bar q^\clad_-]\,.
\end{align}
\end{subequations}
\subsubsection{Interface Conditions}
The interface conditions listed in \Cref{tab:matching gauge-fixed special} yield a set of linear equations,
\begin{equation}
	\label{eq:fibers interface conditions}
	\mathbf M \mathbf N \bs q = 0\,,
\end{equation}
where $\bs q$ is the column vector
\begin{equation}
	\bs q = (q^\core_t, q^\core_\parallel, q^\core_+, q^\core_-, q^\clad_t, q^\clad_\parallel, q^\clad_+, q^\clad_- )^\transpose\,,
\end{equation}
and the matrices $\mathbf M$ and $\mathbf N$ are given by
\begin{widetext}
\begin{align}
	\mathbf M &= \begin{bsmallmatrix}
		1	&	0	&	0	&	0	&	-1	&	0	&	0	&	0\\
		0	&	1	&	0	&	0	&	0	&	-1	&	0	&	0\\
		0	&	0	&	- U \Jj_+	&	0	&	0	&	0	&	W \Kk_+	&	0\\
		0	&	0	&	0 	&	+ U \Jj_-	&	0	&	0	&	0	&	W \Kk_- \\
		- n_1^2 U^2 \Jj	&	0	&	\frac{+i n_1^2 \rho \omega}{\sqrt 2} U \Jj_+ & \frac{-i n_1^2 \rho \omega}{\sqrt 2} U \Jj_-
		& n_2^2 W^2 \Kk	&	0	&	\frac{- i n_2^2 \rho \omega}{\sqrt 2} W \Kk_+ & \frac{- i n_2^2 \rho \omega}{\sqrt 2} W \Kk_-\\
		0	&	0	&	- i U/\sqrt{2} & - i U/\sqrt{2} &
		0	&	0	&	- i W/\sqrt{2} & + i W/\sqrt{2} \\
		0	&	U^2 \Jj	&	\frac{+i \rho \beta}{\sqrt 2} U \Jj_+ & \frac{-i \rho \beta}{\sqrt 2} U \Jj_-
		& 0	&	- W^2 \Kk	&	\frac{- i \rho \beta}{\sqrt 2} W \Kk_+ & \frac{- i \rho \beta}{\sqrt 2} W \Kk_-\\
		i n_1^2 \rho \omega &	i \beta \rho	&	U/\sqrt 2	&	-U/\sqrt 2 &
		- i n_2^2 \rho\omega&	-i \beta \rho	&	W/\sqrt 2	&	W/\sqrt 2
	\end{bsmallmatrix}\,,
	\\
	\mathbf N &= \diag(
		\underbrace{J_m(U), \cdots, J_m(U)}_\text{four times},
		\underbrace{K_m(W), \cdots, K_m(W)}_\text{four times}
	)\,.
\end{align}
\end{widetext}
Here, the following abbreviations were used:
\begin{align}
	\Jj &= \frac{J_m'(U)}{U J_m(U)}\,, &
	\Kk &= \frac{K_m'(W)}{W K_m(W)}\,, \\
	\Jj_\pm &= \Jj \mp m/U^2\,, &
	\Kk_\pm &= \Kk \mp m/W^2\,.
\end{align}
The condition that \cref{eq:fibers interface conditions} admits a non-trivial solution is that the determinant of the matrix $\mathbf M \mathbf N$ vanishes.
One finds
\begin{equation}
	\label{eq:fiber determinant product}
	\det(\mathbf M \mathbf N) = i \mathscr D_1 \mathscr D_2^2\,,
\end{equation}
where
\begin{align}
	\label{eq:fiber determinant 1}
	\begin{split}
		\mathscr D_1
		&= J_m(U)^2 K_m(W)^2
		\\&\quad\times
		\left[
			(\Jj + \Kk)(n_1^2 \Jj + n_2^2 \Kk) - \tilde m^2
		\right]\,,
	\end{split}
	\\
	\label{eq:fiber determinant 2}
	\mathscr D_2
		&= U W J_m(U) K_m(W) \left[
			U^2 \Jj - W^2 \Kk
		\right]\,,
	\\
	\label{eq:fiber determinant m}
	\tilde m
		&= m (\beta / \omega) (U^{-2} + W^{-2})\,.
\end{align}
$\mathscr D_1 = 0$ corresponds to the dispersion relation of Maxwell modes \cite{Liu:2005}, while the modes satisfying $\mathscr D_2 = 0$ correspond to gauge and ghost modes (as shown below).
For given values of the propagation constant $\beta$ and azimuthal mode index $m$, the equations $\mathscr D_1 = 0$ and $\mathscr D_2 = 0$ admit a finite number of solutions for the frequency $\omega$, which are indexed by a radial mode index $\kappa$. In the following, the corresponding solutions will be denoted by
\begin{align*}
	&A_{\phys \beta, m, \kappa}\,,
	&
	&A_{\gauge \beta, m, \kappa}\,,
	&
	&A_{\ghost \beta, m, \kappa}\,,
\end{align*}
where, $A_{\phys \beta, m, \kappa}$ denotes the physical mode whose frequency $\omega$ is the $\kappa$-th solution $\omega = \omega_\kappa$ to the dispersion relation $\mathscr D_1 = 0$ for given values of the propagation constant $\beta$ and azimuthal mode index $m$. The nomenclature of gauge modes $A_{\gauge \beta, m, \kappa}$ (satisfying the gauge condition but having vanishing field strength) and ghost modes $A_{\ghost \beta, m, \kappa}$ (violating the gauge condition) is similar: the radial mode index $\kappa$ iterates over the set of solutions to $\mathscr D_2 = 0$ for given values of $\beta$ and $m$ (see below for details).

\begin{figure*}[t]
	\centering
	\begin{subfigure}[]{\columnwidth}\centering
		\includegraphics[width=\columnwidth]{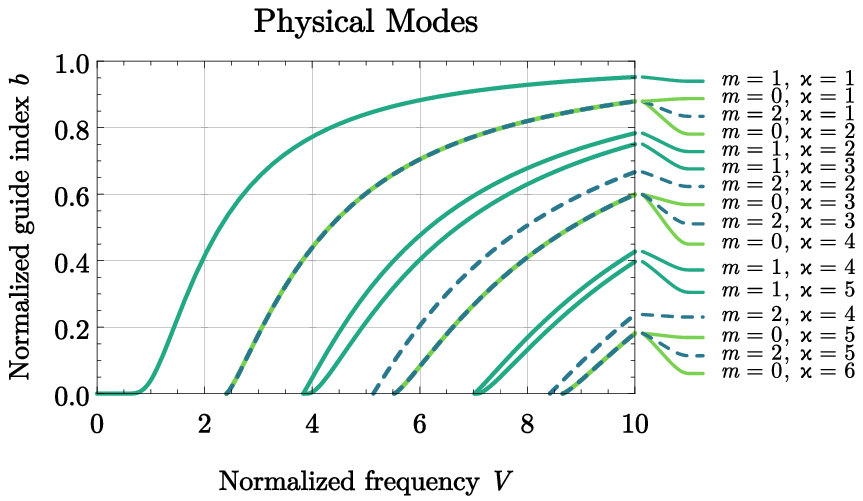}
	\end{subfigure}
	\hspace{1em}
	\begin{subfigure}[]{\columnwidth}\centering
		\includegraphics[width=\columnwidth]{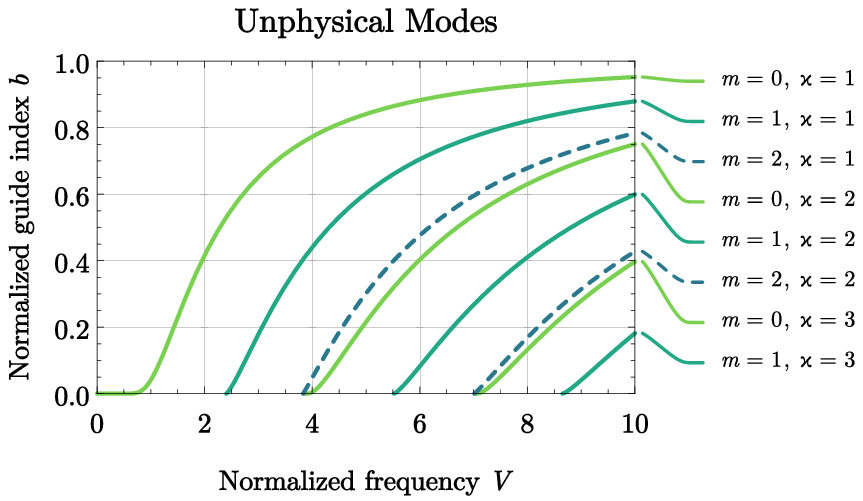}
	\end{subfigure}
	\caption{Mode diagrams of the physical modes (left figure) and unphysical modes (right figure), showing the dependence of the normalized mode index $b = (\beta^2/\omega^2 - n_2^2)/(n_1^2 - n_2^2)$ on the normalized frequency $V = \rho \omega \sqrt{n_1^2 - n_2^2}$. Lines of constant $\beta$ are almost vertical. For illustration, the refractive indices in the core and cladding are set to $n_1 = 1.4712$ and $n_2 = 1.4659$, respectively.}
	\label{fig:mode diagrams}
\end{figure*}

To analyze the dispersion relations, it is customary to define the normalized guide index $b$ and the normalized frequency $V$ as
\begin{align}
	\label{eq:fibers b V definitions}
	b &= \frac{\bar n^2 - n_2^2}{n_1^2 - n_2^2}\,,
	&
	V &= \rho \omega \sqrt{n_1^2 - n_2^2}\,,
\end{align}
where $\bar n = |\beta/\omega|$ is the effective refractive index. Note that $b$ acts as an interpolation parameter expressing the squared effective refractive index as a convex combination of $n_1^2$ and $n_2^2$: $\bar n^2 = b n_1^2 + (1-b) n_2^2$, and $V$ is expressible in terms of $U$ and $W$ defined in \cref{eq:fibers U W} as $V^2 = U^2 + W^2$.
\Cref{fig:mode diagrams} shows the dependence of the normalized guide index $b$ on the normalized frequency $V$ for the physical modes, as well as the gauge and ghost modes. Here, the refractive indices were chosen to be $n_1 = 1.4712$ and $n_2 = 1.4659$, corresponding to typical commercially available single-mode fibers with a core radius of $\rho = 4.1\,\text{µm}$ operated at vacuum wavelengths of $\lambda = 1\,550\,\text{nm}$.
For normalized frequencies $V$ below the threshold $V_* \approx 2.4 $ the fiber supports only a single physical mode (with $m = 1$) and single pair of gauge and ghost modes (with $m = 0$). For larger values of $V$, the fiber supports multiple modes.

\subsubsection{Physical Modes}
\label{s:fiber modes physical}

Physical modes satisfy the dispersion relation
\begin{equation}
	\label{eq:dispersion physical}
	(\Jj + \Kk)(n_1^2 \Jj + n_2^2 \Kk) = \tilde m^2\,,
\end{equation}
with $\tilde m$ as defined in \cref{eq:fiber determinant m}.
Arranging the coefficients $q^\core_i$, $q^\clad_i$, $p^\core_i$ and $p^\clad_i$ in the form
\begin{subequations}
\label{eq:p q matrices}
\begin{align}
	q
		&= \begin{bmatrix}
			\frac{q^\core_t}{J_m(U)}			&
			\frac{q^\core_\parallel}{J_m(U)}	&
			\frac{q^\core_+}{J_m(U)}			&
			\frac{q^\core_-}{J_m(U)}
			\\
			\frac{q^\clad_t}{K_m(W)}			&
			\frac{q^\clad_\parallel}{K_m(W)}	&
			\frac{q^\clad_+}{K_m(W)}			&
			\frac{q^\clad_-}{K_m(W)}
		\end{bmatrix}\,,
	\\
	p
		&= \begin{bmatrix}
			\frac{p^\core_t}{J_m(U)}			&
			\frac{p^\core_\parallel}{J_m(U)}	&
			\frac{p^\core_+}{J_m(U)}			&
			\frac{p^\core_-}{J_m(U)}
			\\
			\frac{p^\clad_t}{K_m(W)}			&
			\frac{p^\clad_\parallel}{K_m(W)}	&
			\frac{p^\clad_+}{K_m(W)}			&
			\frac{p^\clad_-}{K_m(W)}
		\end{bmatrix}\,,
\end{align}
\end{subequations}
the coefficients of the physical modes are
\begin{subequations}
\begin{align}
	\label{eq:mode physical q}
	\begin{split}
		&q_{\phys} = N_{\beta,m,\kappa}^{-1}\\
		&\times
		\begin{bmatrix}
			1 &
			0 &
			\frac{- i \rho \omega}{\sqrt 2 U} [n_1^2 + \frac{\tilde m \beta/\omega}{\Jj + \Kk}] &
			\frac{+ i \rho \omega}{\sqrt 2 U} [n_1^2 - \frac{\tilde m \beta/\omega}{\Jj + \Kk}] \\
			1 &
			0 &
			\frac{+ i \rho \omega}{\sqrt 2 W} [n_2^2 + \frac{\tilde m \beta/\omega}{\Jj + \Kk}] &
			\frac{+ i \rho \omega}{\sqrt 2 W} [n_2^2 - \frac{\tilde m \beta/\omega}{\Jj + \Kk}]
		\end{bmatrix}\,,
	\end{split}
	\\
	\label{eq:mode physical p}
	\begin{split}
		&p_{\phys} = N_{\beta,m,\kappa}^{-1}\\
		&\times
		\begin{bmatrix}
			0 &
			i n_1^2 \beta &
			\frac{- n_1^2 \rho \beta^2}{\sqrt 2 U} [1 + \frac{\tilde m \omega/\beta}{\Jj + \Kk}] &
			\frac{+ n_1^2 \rho \beta^2}{\sqrt 2 U} [1 - \frac{\tilde m \omega/\beta}{\Jj + \Kk}]
			\\
			0 &
			i n_2^2 \beta &
			\frac{+ n_2^2 \rho \beta^2}{\sqrt 2 W} [1 + \frac{\tilde m \omega/\beta}{\Jj + \Kk}] &
			\frac{+ n_2^2 \rho \beta^2}{\sqrt 2 W} [1 - \frac{\tilde m \omega/\beta}{\Jj + \Kk}]
		\end{bmatrix}\,,
	\end{split}
\end{align}
\end{subequations}
where $N_{\beta,m,\kappa}$ is a normalization factor.
Using \cref{eq:fiber gauge function}, one then verifies that the gauge condition is met,
\begin{equation}
	\chi_\phys = 0\,,
\end{equation}
so that Maxwell’s equations are satisfied.

The physical modes with $m = 0$ are either transverse-electric ($\text{TE}$) or transverse-magnetic ($\text{TM}$), while the modes with $m \neq 0$ are hybrid modes commonly classified as either $\text{HE}$ or $\text{EH}$ \cite{Liu:2005}.
For the purposes of the present considerations, however, it suffices to index the physical modes with identical propagation constant $\beta$ and azimuthal mode index $m$ by a radial mode index $\kappa$.

\subsubsection{Gauge Modes}
\label{s:fiber modes gauge}

Gauge modes satisfy the dispersion relation
\begin{equation}
	\label{eq:dispersion gauge}
	U^2 \Jj = W^2 \Kk\,,
\end{equation}
and their coefficients, arranged in the form \eqref{eq:p q matrices}, are
\begin{subequations}
\begin{align}
	\label{eq:mode gauge q}
	q_\gauge &=
	\frac{1}{N'_{\beta, m,\kappa}}
	\begin{bmatrix}
		- i \omega &
		+ i \beta &
		\frac{- U}{\sqrt 2 \rho} &
		\frac{+ U}{\sqrt 2 \rho} \\
		- i \omega &
		+ i \beta &
		\frac{- W}{\sqrt 2 \rho} &
		\frac{- W}{\sqrt 2 \rho}
	\end{bmatrix}\,,
	\\
	\label{eq:mode gauge p}
	p_{\gauge} &= 0\,,
\end{align}
\end{subequations}
where $N'_{\beta, m,\kappa}$ is a normalization factor.
The corresponding field $\emA_\gauge$ is a gradient
\begin{align}
	A_{\gauge \beta,m,\kappa}
		&= \dd \lambda_{\beta,m,\kappa}\,,
	\\
	\shortintertext{where}
	\label{eq:fiber gauge mode}
	\lambda_{\beta,m,\kappa}
		&= f_0(r) e^{i(\beta z + m \theta - \omega t)}\,,
		\\
	f_0(r) &=
		\begin{dcases}
			J_m(U r/\rho)/J_m(U)	&	r < \rho\,,\\
			K_m(W r/\rho)/K_m(W)	&	r > \rho\,.
		\end{dcases}
\end{align}
Using \cref{eq:dispersion gauge} one verifies that $\dd \lambda_{\beta,m,\kappa}$ is continuous, and using \cref{eq:fiber gauge function} one verifies that the gauge condition is satisfied:
\begin{align}
	\chi_\gauge = 0\,.
\end{align}
The momentum density $\emP_{\gauge}$, defined in \cref{eq:fiber momentum}, vanishes because the first term in $\t \emP{^a} = \t\emG{^a^0} - \chi \t\gopt{^a^0}$, is zero as $\emA$ is exact and hence closed, and the second term is zero because the gauge-condition is satisfied.
As a consequence, the gauge modes have vanishing Klein--Gordon norm.

\subsubsection{Ghost Modes}
\label{s:fiber modes ghost}

The ghost modes satisfy the same dispersion relation as the gauge modes:
\begin{equation}
	\label{eq:dispersion ghost}
	U^2 \Jj = W^2 \Kk\,,
\end{equation}
and their coefficients, arranged in the form \eqref{eq:p q matrices}, are
\begin{subequations}
\begin{align}
	\label{eq:mode ghost q}
	q_{\ghost} &=
	\frac{1}{N'_{\beta, m,\kappa}}
	\begin{bmatrix}
		+ i \omega &
		+ i \beta &
		\frac{+ U}{\sqrt 2 \rho} &
		\frac{- U}{\sqrt 2 \rho} \\
		+ i \omega &
		+ i \beta &
		\frac{+ W}{\sqrt 2 \rho} &
		\frac{+ W}{\sqrt 2 \rho}
	\end{bmatrix}\,,
	\\
	\label{eq:mode ghost p}
	p_{\ghost} &=
	\frac{1}{N'_{\beta, m,\kappa}}
	\begin{bmatrix}
		- 2 n_1^2 \beta^2 &
		+ 2 n_1^2 \beta \omega &
		0 &
		0 \\
		- 2 n_2^2 \beta^2 &
		+ 2 n_2^2 \beta \omega &
		0 &
		0
	\end{bmatrix}\,,
\end{align}
\end{subequations}
where, for later convenience, the normalization factor $N'_{\beta, m,\kappa}$ was chosen to be the same as for the gauge modes.
Using \cref{eq:fiber gauge function}, one finds the gauge-violation to be
\begin{equation}
	\label{eq:fiber ghost gauge violation}
	\chi_\ghost = (2i \beta^2/\omega) \t\emA{_\ghost_t}\,.
\end{equation}
One readily verifies that the Klein--Gordon norm of these modes vanishes, as was the case for the gauge modes.

\subsubsection{Klein--Gordon Product}
\label{s:fiber KG product}

The general formula for the Klein--Gordon product, given in \cref{eq:KG product EM general}, yields
\begin{equation}
	\label{eq:fiber KG product}
	\kgbracket{\emA,\emP | \emA', \emP'}
		= i 
		\int_{-\infty}^{+\infty} \hspace{-1.5em} \dd z
		\int_0^{2\pi} \hspace{-1em} \dd \theta
		\int_0^\infty \hspace{-1em} \dd r\, r\,
		\{
			\t*\emA{^*_\mu} \t*\Pi{^\prime^\mu}
			- \t*\emA{^\prime_\mu} \t*\Pi{^*^\mu}
		\}\,.
\end{equation}
As the mode solutions of the form given in \cref{eq:potential decomposition} are not square-integrable, their Klein--Gordon products do not converge.
However, as explained in \Cref{app:KG products wave-packets}, their products are meaningful as integral kernels for the Klein--Gordon product of wave packets.
Factoring the fields $\emA$ and $\emP$ of the modes as
\begin{equation}
	\label{eq:fiber mode factorisation A P}
	\begin{Bmatrix}
		\emA_{I,\beta,m,\kappa}\\
		\emP_{I,\beta,m,\kappa}
	\end{Bmatrix}
	=
	\begin{Bmatrix}
		a_{I,\beta,m,\kappa}(r)\\
		\pi_{I,\beta,m,\kappa}(r)
	\end{Bmatrix}
	e^{i(\beta z + m \theta - \omega t)}\,,
\end{equation}
see \cref{eq:potential decomposition,eq:momentum decomposition}, with $I$ ranging over the labels “\phys”, “\gauge”, and “\ghost”, $\omega = \omega(I,\beta,m,\kappa)$ and $\omega' = \omega'(I', \beta',m',\kappa')$, one obtains
\begin{multline}
	\label{eq:KG product modes general}
		\kgbracket{A_{I,\beta,m,\kappa} | A_{I',\beta',m',\kappa'}}
			= i (2 \pi)^2 \delta_{m, m'} \delta(\beta - \beta') \phantom{\,,}\\
			\times e^{i(\omega - \omega') t} \int_0^\infty \dd r \, r \{
				\t a{^*_\mu} \t \pi{^\prime^\mu}
				- \t a{^\prime_\mu} \t \pi{^*^\mu}
			\} \,,
\end{multline}
where the indices $(I,\beta,m,\kappa)$ of $a$ and $\pi$, as well as the indices $(I',\beta',m',\kappa')$ of $a'$ and $\pi'$ have been suppressed for brevity.

As shown in \Cref{app:orthogonality fiber modes}, modes of different frequencies are orthogonal.
This implies that
(i) the product given in \cref{eq:KG product modes general} is time-independent,
(ii) modes satisfying different dispersion relations are orthogonal, and
(iii) modes satisfying identical dispersion relations but having different radial mode indices $\kappa$ are orthogonal.
Upon normalization of the modes (see \Cref{app:fiber normalization} for details), the only non-trivial Klein--Gordon products of the mode functions are found to be
\begin{subequations}
\label{eq:fiber KG norms}
\begin{align}
	\label{eq:fiber KG norm phys}
	\begin{split}
		&\kgbracket{\emA_{\phys \beta,m,\kappa} | \emA_{\phys \beta',m',\kappa'}}\\
		&\hspace{10em}= \delta_{m,m'} \delta_{\kappa,\kappa'} \delta(\beta - \beta')\,,
	\end{split}
	\\
	\label{eq:fiber KG prod gauge ghost}
	\begin{split}
		&\kgbracket{\emA_{\gauge \beta,m,\kappa} | \emA_{\ghost \beta',m',\kappa'}}\\
		&\hspace{10em}= \delta_{m,m'} \delta_{\kappa,\kappa'} \delta(\beta - \beta')\,.
	\end{split}
\end{align}
\end{subequations}

As we are concerned only with mode solutions of the field equations which have the form considered so far (we ignore, for example, electromagnetic fields propagating mainly outside the optical fibers), we restrict all further discussion to the closure of the space of functions spanned by the modes solutions obtained above.
Accordingly, we consider fields of the form
\begin{multline}
	\label{eq:completeness relation}
		\emA
		= \int_{\mathbf R}\! \dd \beta \sum_{I} \sum_{m \in \mathbf Z} \sum_{\kappa \in K}\\
		\left\{
			a^{(+)}_{I,\beta,m,\kappa} A_{I,\beta, m, \kappa,}
			+ a^{(-)}_{I,\beta,m,\kappa} A^*_{I,\beta, m, \kappa}
		\right\}\,,
\end{multline}
where $I$ ranges over the labels “\phys”, “\gauge”, and “\ghost”, and for each value of $(I,\beta,m)$ the index range $K \equiv K_{I,\beta,m}$ of the radial index $\kappa$ is a finite set.

The first part in \cref{eq:completeness relation}, all of whose terms have a time dependence of the form $e^{- i \omega t}$ with $\omega > 0$, is referred to as the \emph{positive frequency part} of $A$, while the second part is called the \emph{negative frequency part}.
A positive-frequency wave-packet, i.e., a solution of the form given in \cref{eq:completeness relation} with $a^{(-)}_{I,\beta,m,\kappa} \equiv 0$ then has Klein--Gordon norm
\begin{multline}
	\kgbracket{A | A}
		= \int_{\mathbf R} \dd \beta \sum_{m \in \mathbf Z} \sum_{\kappa \in K}
		\{
			|a^{(+)}_{\phys \beta, m, \kappa}|^2 \hspace{1em}\\
			+ (a^{(+)*}_{\ghost \beta, m, \kappa} a^{(+)}_{\gauge \beta, m, \kappa} + \cc)
		\}\,,
\end{multline}
where “\cc” denotes the complex conjugate of the preceding term.

\subsubsection{Quantization}

The quantized theory is now obtained by passing to quantum field operators $\qmA$ and $\qmP$, satisfying the equations of motion as well as the Heisenberg equal-time commutation relation
\begin{align}
	\label{eq:commutator Heisenberg}
	[\t\qmA{_a}(t, \bs x), \t\qmP{^b}(t, \bs y)]
		= i\hbar\, \t*\delta{_a^b} \delta(\bs x - \bs y)\,.
\end{align}
For any positive-frequency solution $\psi$ of the field equations i.e., a solution of the form \eqref{eq:completeness relation} with $a^{(-)}_{I,\beta,m,\kappa} \equiv 0$, define the ladder operators
\begin{align}
	\hat a(\psi) &= + \frac{1}{\sqrt\hbar} \kgbracket{\psi | \qmA}\,,
	&
	\hat a^\dagger(\psi) &= - \frac{1}{\sqrt\hbar} \kgbracket{\psi^* | \qmA}\,,
\end{align}
then \cref{eq:commutator Heisenberg} implies
\begin{equation}
	\label{eq:commutator ladder general}
	[\hat a(\varphi), \hat a^\dagger(\psi)]
		= \kgbracket{\varphi | \psi}\,.
\end{equation}
Defining
\begin{subequations}
\label{eq:waveguide ladder operators}
\begin{align}
	\hat a_{\beta,m,\kappa}
		&= \hat a(A_{\phys \beta,m,\kappa})\,,
	\\
	\hat b_{\beta,m,\kappa}
		&= \hat a(A_{\gauge \beta,m,\kappa})\,,
	\\
	\hat c_{\beta,m,\kappa}
		&= \hat a(A_{\ghost \beta,m,\kappa})\,,
\end{align}
\end{subequations}
it follows from \cref{eq:fiber KG norms,eq:commutator ladder general} that the only non-vanishing commutators of these ladder operators are
\begin{equation}
	\label{eq:fiber commutators}
	\begin{split}
		[\hat a_{\beta,m,\kappa}, \hat a^\dagger_{\beta',m',\kappa'}]
		&= [\hat b_{\beta,m,\kappa}, \hat c^\dagger_{\beta',m',\kappa}]\\
		&= \delta_{m,m'} \delta_{\kappa,\kappa'} \delta(\beta - \beta')\,.
	\end{split}
\end{equation}
As in \eqref{eq:completeness relation}, the field operator can be decomposed as
\begin{equation}
\begin{split}
	\label{eq:fiber field op}
	\qmA
		= \sqrt\hbar \sum_{m,\kappa} \int\!\dd\beta \bigg\{
			&\hat a_{\beta,m,\kappa} A_{\phys\beta,m,\kappa}\\
			&+ \hat c_{\beta,m,\kappa} A_{\gauge\beta,m,\kappa}\\
			&+ \hat b_{\beta,m,\kappa} A_{\ghost\beta,m,\kappa}
		\bigg\} + \text{h.c.} \,,
	\end{split}
\end{equation}
(“h.c.” denotes Hermitian conjugation)
where the ladder operators of gauge excitations multiply the ghost-mode functions (and vice versa).
This is because both gauge and ghost modes have vanishing Klein--Gordon norm, but have a non-zero mutual Klein--Gordon product.

From \cref{eq:fiber commutators,eq:fiber field op} one deduces the relations
\begin{subequations}
	\label{eq:fiber commutators A ladder}
	\begin{align}
		\label{eq:fiber commutators A ladder a}
		[\qmA, \hat a_{\beta,m,\kappa}^\dagger] &= A_{\phys \beta,m,\kappa}\,,\\
		\label{eq:fiber commutators A ladder b}
		[\qmA, \hat b_{\beta,m,\kappa}^\dagger] &= A_{\gauge \beta,m,\kappa}\,,\\
		\label{eq:fiber commutators A ladder c}
		[\qmA, \hat c_{\beta,m,\kappa}^\dagger] &= A_{\ghost \beta,m,\kappa}\,,
	\end{align}
\end{subequations}
expressing the fact that the operators $\hat a_{\beta,m,\kappa}^\dagger$, $\hat b_{\beta,m,\kappa}^\dagger$ and $\hat c_{\beta,m,\kappa}^\dagger$ generate physical, gauge, and ghost excitations, respectively.

The operator corresponding to the gauge function $\chi$ is
\begin{equation}
	\label{eq:fiber gauge function operator}
	\hat \chi
		= \sqrt\hbar \sum_{m,\kappa} \int\!\dd\beta \left\{
			\hat b_{\beta,m,\kappa} \chi_{\ghost \beta,m,\kappa}
			+ \text{h.c.}
		\right\}\,,
\end{equation}
where $\chi_{\ghost \beta,m,\kappa}$ is the gauge violation of the ghost modes given in \cref{eq:fiber ghost gauge violation}.
So far, the unphysical ghost-excitations have not yet been eliminated from the theory. This is accomplished using the Gupta--Bleuler condition, which requires all physical states $\ket\Phi$ to satisfy%
\begin{equation}
	\label{eq:GB condition}
	\hat \chi^{(+)} \ket\Phi = 0\,,
\end{equation}
where $\hat \chi^{(+)}$ denotes the positive-frequency part of $\hat \chi$, which is given by the part written explicitly in \cref{eq:fiber gauge function operator}, i.e., without the “h.c.”-term.
\Cref{eq:GB condition} is equivalent to the condition%
\begin{equation}
	\label{eq:GB condition 2}
	\hat b_{\beta,m,\kappa} \ket \Phi = 0\,,
\end{equation}
for all $\beta$ and $m$ and $\kappa$. This condition requires all physical states to be free of ghost-excitations generated by $\hat c_{\beta,m,\kappa}^\dagger$, but allows for arbitrary gauge-excitations generated by $\hat b_{\beta,m,\kappa}^\dagger$, cf.\ \cref{eq:fiber commutators}.

A direct consequence of \cref{eq:GB condition} is that the gauge condition is satisfied in the sense of “matrix elements” of physical states $\ket\Phi$ and $\ket\Psi$ in the sense that $\braket{\Phi | \hat\chi | \Psi} = 0$, since
\begin{equation}
	\braket{\Phi | \hat\chi | \Psi}
		= \braket{\Phi | \hat \chi^{(+)} | \Psi}
		+ \braket{\Phi | \hat \chi^{(-)} | \Psi}
		= 0\,,
\end{equation}
as $\chi^{(+)} \ket\Psi = 0$ and $\bra\Phi \hat \chi^{(-)} = [\chi^{(+)} \ket\Phi]^\dagger = 0$.

\subsubsection{Gauge Invariance}

To analyze the action of the operators $\hat b^\dagger_{\beta,m,\kappa}$, generating gauge-excitations, let $\ket{\Psi'}$ and $\ket{\Psi''}$ be two physical states (satisfying the Gupta--Bleuler condition), and consider the linear combination%
\begin{equation}
	\ket{\Psi}
		= \ket{\Psi'} + \hat b^\dagger_{\beta,m,\kappa} \ket{\Psi''}\,.
\end{equation}
If $\ket\Phi$ is any other physical state, then
\begin{equation}
	\braket{\Phi | \Psi}
		= \braket{\Phi | \Psi'}
		+ \braket{\Phi | \hat b^\dagger_{\beta,m,\kappa} | \Psi''}
		= \braket{\Phi | \Psi'}\,,
\end{equation}
since $\bra\Phi \hat b^\dagger_{\beta,m,\kappa} = [\hat b_{\beta,m,\kappa} \ket \Phi]^\dagger = 0$.
Therefore, gauge-excitations do not contribute to inner products of physical states.
Moreover, for the “matrix element” $\braket{\Phi | \qmA | \Psi}$ one has the gauge-transformation formula
\begin{equation}
	\braket{\Phi | \qmA | \Psi}
		= \braket{\Phi | \qmA | \Psi'}
		+ \braket{\Phi | \Psi''} \sqrt\hbar\, \dd \lambda_{\beta,m,\kappa}\,,
\end{equation}
with $\lambda_{\beta,m,\kappa}$ as defined in \cref{eq:fiber gauge mode}.
This follows from $[\qmA, \hat b^\dagger_{\beta,m,\kappa}] = \sqrt\hbar A_{\gauge \beta,m,\kappa} = \sqrt\hbar \dd \lambda_{\beta,m,\kappa}$, see \cref{eq:fiber commutators A ladder b}.
As a consequence, all operators $\hat O$ commuting with $\hat b_{\beta,m,\kappa}$ and $\hat b_{\beta,m,\kappa}^\dagger$ are gauge-invariant in the sense that their “matrix elements” $\braket{\Phi | \hat O | \Psi}$ are independent of gauge-excitations of the states $\ket\Phi$ and $\ket\Psi$.
In particular, the operators
\begin{align*}
	&\qmF\,,
	&
	&\qmG\,,
	&
	&\oint_\gamma \qmA\,,
	&
	&\int_M \t\qmA{_a} \t j{^a} \lc\,,
\end{align*}
where $\gamma$ is any loop in space-time, and $\t j{^a}$ is any divergence-free vector field decaying at infinity, are gauge-invariant operators (commuting with the $\hat b_{\beta,m,\kappa}$ operators) and are thus physical observables.

\subsubsection{Wave Packets}

The physical modes considered so far have infinite norm, see \cref{eq:fiber KG norm phys}. This is remedied by considering wave packets instead.
For example, the wave packet
\begin{equation}
	A_\psi = \int \dd \beta \sum_{m,\kappa} \psi(\beta, m, \kappa) A_{\phys \beta,m,\kappa}
\end{equation}
is normalized whenever
\begin{equation}
	\int \dd \beta \sum_{m,\kappa} |\psi(\beta, m, \kappa)|^2 = 1\,.
\end{equation}
The associated ladder operator $\hat a_\psi^\dagger$ is then given by
\begin{equation}
	\hat a_\psi^\dagger
		= \hat a^\dagger(A_\psi)
		= \int \dd \beta \sum_{m,\kappa} \psi(\beta, m, \kappa) \hat a^\dagger_{\beta,m,\kappa}\,,
\end{equation}
and satisfies
\begin{equation}
	[\t{\hat a}{_\psi}, \t*{\hat a}{_\psi^\dagger}] = 1\,,
\end{equation}
so the corresponding singe-photon state $\hat a_\psi^\dagger \ket 0$ is properly normalized.
At this stage, the standard quantum optics notions of multi-photon states, coherent states, etc.\ can be used.

\subsection{Optical Fibers in Weak Gravitational Fields}
\label{s:fibers in curved space}

In this section, the previous calculations on optical fibers in flat space-time are generalized to describe optical fibers placed horizontally in a homogeneous gravitational field.
The resulting model is then applied to Mach--Zehnder interferometers in Earth’s gravitational field.

\subsubsection{Linearized Parametrized-Post-Newtonian Metrics}

The linearized Einstein field equations for static mass distributions lead to the post-Newtonian metric
\begin{equation}
	\label{eq:metric PN}
	g_\text{PN}
		= - (1 + 2 \gravP) \dd t^2 + (1 - 2 \gravP) \t\delta{_i_j} \t{\dd x}{^i} \t{\dd x}{^j}\,,
\end{equation}
see, e.g., Ref.~\cite{Wald:1984}.
For experimental tests of general relativity, it is customary to consider a wider class of weak-field metrics: the parametrized post-Newtonian (PPN) metrics \cite{2014LRR....17....4W}. In the linearized scheme, the PPN metrics take the form
\begin{equation}
	\label{eq:metric PPN}
	g_\text{PPN}
		= - (1 + 2 \aLPI \gravP) \dd t^2 + (1 - 2 \gamma \gravP) \t\delta{_i_j} \t{\dd x}{^i} \t{\dd x}{^j}\,,
\end{equation}
where deviations of the LPI-parameter $\aLPI$ from $1$ describe violations of local position invariance (LPI) \cite{2014LRR....17....4W}, and the PPN-parameter $\gamma$ quantifies the “strength of spatial curvature per unit rest mass” \cite{2014LRR....17....4W}. In general relativity, both parameters are equal to $1$, while Nordström’s theory, for example, corresponds to $\aLPI = 1$ and $\gamma = -1$.

For experiments such as the one proposed in Ref.~\cite{2017NJPh...19c3028H}, where optical fibers are placed horizontally in Earth’s gravitational field to form an interferometer, the gravitational potential $\gravP$ is constant along each fiber to good precision. While the above calculations of \Cref{s:fibers in flat space} can be generalized to the PN or PPN metrics, (for PN metrics, this was done at the level of the classical electromagnetic field strength $\emF$ and excitation $\emG$ in Ref.~\cite{2018CQGra..35x4001B}), a more direct route is used here.

Assuming a homogeneous field with potential
\begin{equation}
	\label{eq:potential linear}
	\gravP = \gravP_0 + \gravA z\,,
\end{equation}
one can use the coordinate transformation and rescaling of $\gravP$ described in \Cref{app:metric reduction PPN to N} to reduce the metric to the Newtonian form
\begin{equation}
	\label{eq:metric N}
	g_\text{N}
		= - (1 + 2 \gravP) \dd t^2 + \t\delta{_i_j} \t{\dd x}{^i} \t{\dd x}{^j}\,.
\end{equation}
The linear term in \cref{eq:potential linear} gives only small corrections to the mode profiles \cite{2015PhRvD..91f4041B,2018CQGra..35x4001B} and can thus be neglected in a first approximation (for typical optical fibers, the core radius is of the order of micrometers and hence $\gravA \rho \approx \num{1e-22}$ while at Earth’s surface $\varphi_0 \approx \num{7e-10}$). The main influence of the gravitational field on a horizontal optical fiber is thus contained in the constant term $\gravP_0$, which encodes the gravitational redshift between fibers at different altitudes (for which the numerical values of $\gravP_0$ differ).
As shown in the next section, within this approximation, the gravitationally perturbed fiber modes are obtained from the unperturbed ones by simple substitutions.

It is worth noting that the transformation from \cref{eq:metric PPN} to \cref{eq:metric N} shows that experiments insensitive to gravity gradients, i.e., insensitive to space-time curvature, are necessarily insensitive to the PPN parameter $\gamma$ \cite{2015PhRvD..91f4041B}.

\subsubsection{Perturbation of Fiber Modes}

For a linear homogeneous dielectric at rest in the coordinate system considered, the covariant optical metric reads
\begin{equation}
	\gopt = - n^{-2} (1 + 2 \gravP) \dd t^2 + \t\delta{_i_j} \t{\dd x}{^i} \t{\dd x}{^j}\,.
\end{equation}
Since the field equations for a constant refractive index $n$ are formulated entirely in terms of the optical metric $\gopt$, the rescaling $n \to (1 - \gravP) n$ transforms solutions $\t\emA{_a}$ of the field equations in Minkowski space to solutions for a uniform potential. Moreover, since the interface conditions listed in \Cref{tab:matching gauge-fixed general} are independent of the gravitational potential $\gravP$, all continuity conditions remain satisfied.
However, the modes obtained in this manner are not correctly normalized.
The gravitational field produces a rescaling in the space-time volume-form according to $\lc \to (1 + \gravP) \lc$, but leaves the spatial volume-form $\formV$ invariant. According to \cref{eq:form P s-t} and the factorization $\formP = \emP \otimes \formV$, the canonical momentum is thus rescaled by $\emP \to (1 + \gravP) \emP$, and so are the Klein--Gordon products.
To maintain the normalization of the modes, one thus has to rescale the normalization factors $N_{\beta,m,\kappa}$ and $N'_{\beta,m,\kappa}$ of \cref{eq:mode physical q,eq:mode physical p,eq:mode gauge q,eq:mode ghost q,eq:mode ghost p} according to
\begin{align}
	N_{\beta,m,\kappa} &\to (1 + \gravP) N_{\beta,m,\kappa}\,,
	&
	N'_{\beta,m,\kappa} &\to (1 + \gravP) N'_{\beta,m,\kappa}\,.
\end{align}
Finally, since the spatial volume form $\formV$ is unchanged, so are the Heisenberg equal-time commutation relations \eqref{eq:commutator Heisenberg}.
This completes the transition from fiber modes in flat space to those in a homogeneous gravitational potential.

\subsection{Gravitational Phase Shift}
\label{s:gravitational phase shift}

The relation between the propagation constant $\beta$ and the frequency $\omega$ is encoded in the effective refractive index, $\bar n$, via $\beta^2 = \bar n^2 \omega^2$, where $\bar n^2$ can be expressed as a convex combination of the squared refractive index of the core, $n_1^2$, and that of the cladding, $n_2^2$, as $\bar n^2 = b n_1^2 + (1 - b) n_2^2$.
Here, $b$ is the normalized guide index defined in \cref{eq:fibers b V definitions}, which is a function of the normalized frequency $V = \rho \omega \sqrt{n_1^2 - n_2^2}$.
To first order in the gravitational potential $\gravP$, the substitution $n \to (1 - \gravP)n$, mapping solutions in flat space-time to solutions in a homogeneous gravitational field, yields
\begin{equation}
	\bar n^2
		\to (1 - \gravP)^2 \bar n^2 - \frac{\p b}{\p V} (n_1^2 - n_2^2) \gravP\,.
\end{equation}
For weakly guiding fibers, where $(n_1^2 - n_2^2) \ll \bar n^2$, the second term is negligible compared to the first.
With this approximation, one obtains the dispersion relation
\begin{equation}
	\beta^2 = (1 - \gravP)^2 \bar n^2 \omega^2\,,
\end{equation}
in which $\bar n$ is the effective refractive index as calculated in flat space-time.

The frequency $\omega$ arising here is a coordinate-frequency, related to the physical frequency (as measured in the rest-frame of the dielectric) by $\omega_\ph = (1 - \gravP) \omega$, while $\beta_\ph \equiv \beta$ directly corresponds to the spatial norm of phase one-form $\beta \dd z$.
The relationship between $\beta_\ph$ and $\omega_\ph$ then takes the standard form
\begin{equation}
	\label{eq:invariant refractive index}
	\beta_\ph^2 = \bar n^2 \omega_\ph^2\,,
\end{equation}
showing that (in the weakly guiding limit) the physical effective refractive index is unperturbed by the uniform gravitational field.
As a consequence, the gravitational phase shifts in optical fibers are explicable purely in terms of gravitational redshifts.

For a Mach--Zehnder interferometer consisting of optical fibers of length $L$, separated in height by $\Delta z$ with $\Delta z$ much smaller than the local radius of space-time curvature, this directly gives rise to the gravitational phase shift formula \cite{2017NJPh...19c3028H,2018CQGra..35x4001B}
\begin{equation}
	\label{eq:gravitational phase shift}
	\Delta \psi
		= - \bar n \omega \gravA L \Delta z\,,
\end{equation}
where $\gravA$ is the local gravitational acceleration. In \cref{eq:gravitational phase shift} the distinction between $\omega$ and $\omega_\ph$ is irrelevant to leading order.

\subsection{Gravitational Redshift}
\label{s:redshift mode transformation}

For identical optical fibers placed at different gravitational potentials $\gravP'$ and $\gravP''$, the above quantization scheme yields mode operators $\hat\alpha'(\beta')$ and $\hat\alpha''(\beta'')$ (where further indices pertaining to the azimuthal and radial mode indices are suppressed for brevity), which are parametrized by their \emph{proper propagation constants} $\beta' = \bar n \omega'$ and $\beta'' = \bar n \omega''$ as measured by local observers in the gravitational potentials $\gravP'$ and $\gravP''$, respectively. Here, $\bar n$ is the effective refractive index as in flat space-time, see \cref{eq:invariant refractive index}.
The relation between the two proper frequencies $\omega'$ and $\omega''$ is expressible in terms of the “Killing frequency” $\omega$ as
\begin{equation}
	\label{eq:redshift omega linear}
	\omega = (1 + \gravP') \omega' = (1 + \gravP'') \omega''\,.
\end{equation}
It is advantageous to define a “Killing propagation constant” $\beta = \bar n \omega$ such that an analogous equation holds for the propagation constants:
\begin{equation}
	\label{eq:redshift beta linear}
	\beta = (1 + \gravP') \beta' = (1 + \gravP'') \beta''\,.
\end{equation}
To describe the gravitational redshift of quantum states, one must relate the ladder operators $\hat\alpha'(\beta')$ and $\hat\alpha''(\beta'')$. As was shown in Ref.~\cite{2021arXiv210900728B}, this relationship is given by
\begin{equation}
	\label{eq:redshift mode relations BS}
	\hat\alpha'(\beta') = \sqrt{1 + \gravP' - \gravP''} \hat\alpha''(\beta'')\,,
\end{equation}
where the square-root factor is necessary for compatibility with the commutation relations
\begin{subequations}
\begin{align}
	[\hat\alpha'(\beta'_1), \hat\alpha'(\beta'_2)^\dagger] &= \delta(\beta'_1 - \beta'_2)\,,
	\\
	[\hat\alpha''(\beta''_1), \hat\alpha''(\beta''_2)^\dagger] &= \delta(\beta''_1 - \beta''_2)\,.
\end{align}
\end{subequations}
A more symmetric description is obtained, however, by introducing new ladder operators parametrized by the Killing propagation constant $\beta$:
\begin{subequations}
\label{eq:redshift mode operators}
\begin{align}
	\hat a'(\beta) &= \sqrt{1-\gravP'}\, \hat \alpha'(\beta')\,,
	\\
	\hat a''(\beta) &= \sqrt{1-\gravP''}\, \hat \alpha''(\beta'')\,,
\end{align}
\end{subequations}
satisfying the commutation relations
\begin{equation}
	[\hat a'(\beta_1), \hat a'(\beta_2)^\dagger]
	= [\hat a''(\beta_1), \hat a''(\beta_2)^\dagger]
	= \delta(\beta_1 - \beta_2)\,.
\end{equation}
In terms of these operators, \cref{eq:redshift mode relations BS} takes the simple form
\begin{equation}
	\hat a'(\beta) = \hat a''(\beta)\,.
\end{equation}

As an application of this formalism, consider a single-photon wave packet sent from a region of gravitational potential $\gravP'$ to a region with gravitational potential $\gravP''$. In the Heisenberg picture the state vector remains constant. When the wave packet is decomposed in terms of the $\hat \alpha'(\beta')$ and $\hat \alpha''(\beta'')$ bases, the associated “wave functions” $\psi'$ and $\psi''$, defined by
\begin{equation}
	\begin{split}
		\ket \psi
			&= \int \hspace{-0.25em} \dd \beta'\, \psi'(\beta') \hat \alpha'(\beta')^\dagger \ket 0\\
			&= \int \hspace{-0.25em} \dd \beta''\, \psi''(\beta'') \hat \alpha''(\beta')^\dagger \ket 0\,,
	\end{split}
\end{equation}
are related by the non-trivial transformation
\begin{equation}
	\psi''(\beta'') = \sqrt{1 - \gravP' + \gravP''} \psi'(\beta')\,.
\end{equation}
However, when the mode operators $\hat a'(\beta)$ and $\hat a''(\beta)$ are used, no such transformation rule is necessary:
\begin{equation}
	\begin{split}
		\ket \psi
			&= \int \hspace{-0.25em} \dd \beta\, \psi(\beta) \hat \alpha'(\beta)^\dagger \ket 0\\
			&= \int \hspace{-0.25em} \dd \beta\, \psi(\beta) \hat \alpha''(\beta)^\dagger \ket 0\,.
	\end{split}
\end{equation}

\subsection{Gravitational Photon Interferometry I}

\begin{figure}[b]
	\centering
	\includegraphics[width=\columnwidth]{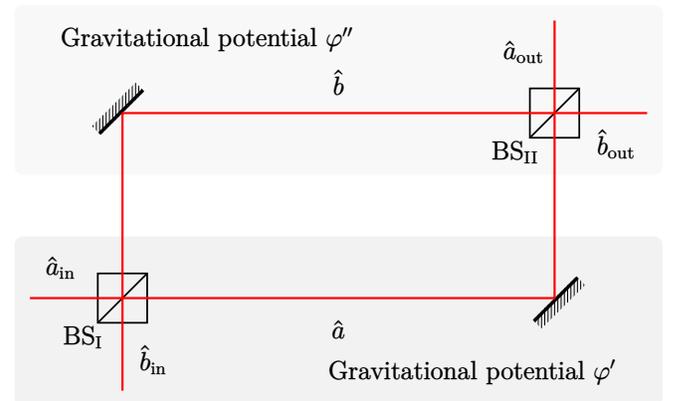}
	\caption{Schematic drawing of a Mach--Zehnder interferometer in a gravitational field, with the two horizontal arms placed at constant gravitational potentials $\gravP'$ and $\gravP''$, each.}
	\label{fig:schematics Mach--Zehnder}
\end{figure}

The quantization procedure described here provides a consistent framework for the description of single-photon and multi-photon interferometry in Earth’s gravitational field, as conceived in Refs.~\cite{2017NJPh...19c3028H,2022PhRvA.106c1701M} and illustrated in \Cref{fig:schematics Mach--Zehnder}.
These references consider Mach--Zehnder interferometers formed by optical fibers placed horizontally in a gravitational field, which are separated in altitude by a height difference $h$. These two horizontal arms are joined by vertical fibers (or free space), which produce identical phase shifts and are thus irrelevant for the interferometric phase.

For sufficiently large separations $h$, the electromagnetic fields in the two fibers can be considered as decoupled, so that the total Hilbert space of states is given by the tensor products of the spaces associated to the individual fibers.

The mode nomenclature of \Cref{fig:schematics Mach--Zehnder} is as follows: $\hat a_\text{in}$, $\hat b_\text{in}$ denote the ladder operators of the input modes, $\hat a_\text{out}$, $\hat b_\text{out}$ pertain to the output modes, while the ladder operators associated to the horizontal interferometer arms at constant potentials $\gravP'$ and $\gravP''$ are denoted by $\hat a$, $\hat b$, respectively.
All of them are taken to be parametrized by invariant Killing propagation constants $\beta$, as described in \Cref{s:redshift mode transformation}.
Moreover, the phases of the all modes at the first beam splitter, and the output modes at the second beam splitter, are set to $- \omega t$, where $\omega$ is the Killing frequency, and $t$ is Killing time.
Finally, the phases of the modes $\hat a$ and $\hat b$ at the second beam splitter are denoted by $\psi'$ and $\psi''$ (they depend on the local gravitational potentials $\gravP'$ and $\gravP''$, respectively).
If both beam splitters are symmetric and lossless, the mode coupling at the beam splitters takes the form
\begin{align}
	\begin{bmatrix}
		\hat a_\beta\\
		\hat b_\beta
	\end{bmatrix}
	&= \frac{1}{\sqrt 2}
	\begin{bmatrix}
		1	&	i\\
		i	&	1
	\end{bmatrix}
	\begin{bmatrix}
		\hat a_\text{in}\\
		\hat b_\text{in}
	\end{bmatrix}\,,
	\\
	\begin{bmatrix}
		\hat a_\text{out}\\
		\hat b_\text{out}
	\end{bmatrix}
	&=
	\frac{1}{\sqrt 2}
	\begin{bmatrix}
		1	&	i\\
		i	&	1
	\end{bmatrix}
	\begin{bmatrix}
		e^{i \psi'}	&	0\\
		0	&	e^{i \psi''}
	\end{bmatrix}
	\begin{bmatrix}
		\hat a_\beta\\
		\hat b_\beta
	\end{bmatrix}\,.
\end{align}
Conversely, for the creation operators, one has
\begin{align}
	\begin{bmatrix}
		\hat a_\text{in}^\dagger\\
		\hat b_\text{in}^\dagger
	\end{bmatrix}
	&= \frac{1}{\sqrt 2}
	\begin{bmatrix}
		1	&	i\\
		i	&	1
	\end{bmatrix}
	\begin{bmatrix}
		\hat a_\beta^\dagger\\
		\hat b_\beta^\dagger
	\end{bmatrix}
	\,,
	\\
	\begin{bmatrix}
		\hat a_\beta^\dagger\\
		\hat b_\beta^\dagger
	\end{bmatrix}
	&=
	\frac{1}{\sqrt 2}
	\begin{bmatrix}
		e^{i \psi'}	&	0\\
		0	&	e^{i \psi''}
	\end{bmatrix}
	\begin{bmatrix}
		1	&	i\\
		i	&	1
	\end{bmatrix}
	\begin{bmatrix}
		\hat a_\text{out}^\dagger\\
		\hat b_\text{out}^\dagger
	\end{bmatrix}
	\,.
\end{align}
All in all, the input operators are related to the output operators by
\begin{equation}
	\label{eq:MZI mode relation}
	\begin{bmatrix}
		\hat a_{\text{in}}^\dagger\\
		\hat b_{\text{in}}^\dagger
	\end{bmatrix}
	=
	\begin{bmatrix}
		\half (e^{i \psi'} - e^{i \psi''})
			& +\ihalf (e^{i \psi'} + e^{i \psi''}) \\
		\ihalf (e^{i \psi'} + e^{i \psi''})
			& - \half (e^{i \psi'} - e^{i \psi''})
	\end{bmatrix}
	\begin{bmatrix}
		\hat a_{\text{out}}^\dagger\\
		\hat b_{\text{out}}^\dagger
	\end{bmatrix}\,,
\end{equation}
where the $\beta$-dependence of all quantities has been suppressed for brevity.

These relations allow the determination of the interferometer output for arbitrary inputs.
For example, consider a coherent input state
\begin{multline}
	\ket{\coherent(\alpha_1, \alpha_2)}_\text{in}
		= e^{-(|\alpha_1|^2 + |\alpha_2|^2)/2}\\
		\times \exp\left(
			\alpha_1 \hat a^\dagger_\text{in}
			+ \alpha_2 \hat b^\dagger_\text{in}
		\right) \ket 0\,,
\end{multline}
for any complex numbers $\alpha_1$ and $\alpha_2$.
Using \cref{eq:MZI mode relation} one finds the output (which equals the input in the Heisenberg picture) to be coherent also when expressed in the out-basis:
\begin{multline}
	\ket{\coherent(\alpha_1', \alpha_2')}_\text{out}
		= e^{-(|\alpha_1'|^2 + |\alpha_2'|^2)/2}\\
		\times \exp\left(
			\alpha_1' \hat a^\dagger_\text{out}
			+ \alpha_2' \hat b^\dagger_\text{out}
		\right) \ket 0\,,
\end{multline}
where the parameters $\alpha_1'$ and $\alpha_2'$ are related to $\alpha_1$ and $\alpha_2$ by the linear transformation
\begin{equation}
	\begin{bmatrix}
		\alpha_1'\\
		\alpha_2'
	\end{bmatrix}
	= \begin{bmatrix}
		\half (e^{i \psi'} - e^{i \psi''})
			& +\ihalf (e^{i \psi'} + e^{i \psi''}) \\
		\ihalf (e^{i \psi'} + e^{i \psi''})
			& - \half (e^{i \psi'} - e^{i \psi''})
	\end{bmatrix}
	\begin{bmatrix}
		\alpha_1\\
		\alpha_2
	\end{bmatrix}\,.
\end{equation}
Further, if a single photon is sent into the $\hat a_\text{in}$ mode, one obtains
\begin{multline}
	\ket{1,0}_\text{in}
		= \half (e^{i \psi'} - e^{i \psi''}) \ket{1,0}_\text{out}\\
		+ \ihalf (e^{i \psi'} + e^{i \psi''}) \ket{0,1}_\text{out}\,,
\end{multline}
so that the probability of finding one photon in the $\hat a_\text{out}$ mode is
\begin{equation}
	\label{eq:single photon p1}
	p_1
		= \half[1 - \cos(\Delta \psi)]\,,
\end{equation}
with the phase shift $\Delta \psi = \psi' - \psi''$.
Alternatively, for a two-photon state, where one photon is sent into each of the inputs, one has
\begin{multline}
	\ket{1,1}_\text{in}
		= \tfrac{i}{2 \sqrt 2} (e^{2 i \psi'} - e^{2 i \psi''})(\ket{2,0}_\text{out} - \ket{0,2}_\text{out})\\
		- \half (e^{2 i \psi'} + e^{2 i \psi''}) \ket{1,1}_\text{out}\,.
\end{multline}
In this case, the probability of finding both photons in the same output mode is thus
\begin{equation}
	\label{eq:two photon p2}
	p_2
		= \half[1 - \cos(2 \Delta \psi)]\,,
\end{equation}
so that for this particular two-photon input, the fringe frequency is doubled compared to the single-photon probability given in \cref{eq:single photon p1}.
More detailed calculations with wave packets of finite bandwidth show that, compared to single-photon interferometry, such two-photon interference experiments allow for a reduction of the height differences necessary to resolve gravity gradients by more than a factor of two \cite{2022PhRvA.106c1701M}.

Note that the description provided here does not require the linear approximation \eqref{eq:potential linear} to hold across the entire interferometer, but only requires the gravitational field to be approximately homogeneous over the region occupied by each interferometer arm separately. Accordingly, the constants $\gravP_0$ and $\gravA$ may differ for the two fibers considered, so that the current model also applies to experiments probing gravity gradients.

\subsection{Gravitational Photon Interferometry II}

\begin{figure}[t]
	\centering
	\includegraphics[width=\columnwidth]{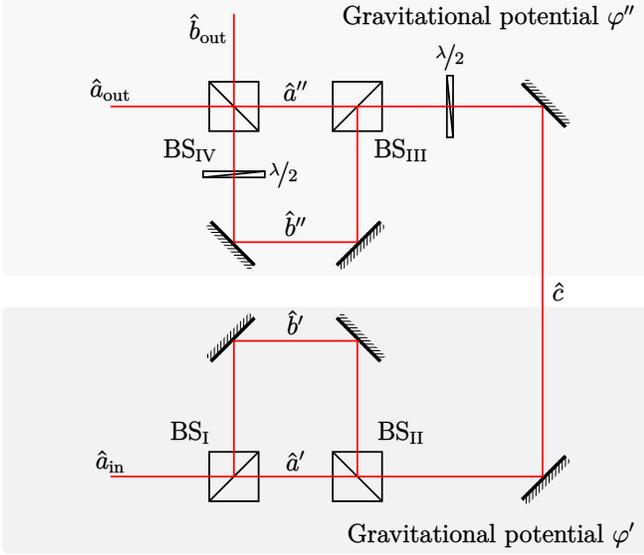}
	\caption{Schematic drawing of a time-bin encoding interferometer with the unbalanced Mach--Zehnder sub-interferometers placed at different gravitational potentials $\gravP'$ and $\gravP''$.}
	\label{fig:schematics TBI}
\end{figure}

An alternative interferometer layout to be used for gravitational quantum optics experiments is sketched in \Cref{fig:schematics TBI} \cite{2012CQGra..29v4011R,2021arXiv211115591M}.

Denoting by $\beta'$ and $\beta''$ the local propagation constants in the lower and upper parts of the interferometer, respectively, and similarly for the delay arm lengths $l'$ and $l''$, one finds
\begin{equation}
	\hat a^\dagger_\text{in}
		= \ihalf (e^{i \beta' l'} + e^{i \beta'' l''}) \hat a^\dagger_\text{out}
		- \half (e^{i \beta' l'} - e^{i \beta'' l''}) \hat a^\dagger_\text{out}\,.
\end{equation}
If both delay lines are taken to be equally long, $l' = l''$, the final result up to a physically irrelevant overall phase is
\begin{equation}
	\hat a^\dagger_\text{in}
		\propto \ihalf (1 + e^{-i (\beta' - \beta'') l'}) \hat a^\dagger_\text{out}
		- \half (1 - e^{-i (\beta' - \beta'') l'}) \hat a^\dagger_\text{out}\,.
\end{equation}
\Cref{eq:redshift beta linear} relates the difference in propagation constant, $\beta' - \beta''$, to the difference in gravitational potential, $\Delta\gravP = \gravP' - \gravP''$, according to $\beta' - \beta'' = \beta' \Delta \gravP$, which leads to
\begin{equation}
	\hat a^\dagger_\text{in}
		\propto \ihalf (1 + e^{-i \beta' l' \Delta \gravP}) \hat a^\dagger_\text{out}
		- \half (1 - e^{-i \beta' l' \Delta \gravP}) \hat a^\dagger_\text{out}\,.
\end{equation}
Accordingly, the output probabilities of finding a photon in mode $\hat a$ or $\hat b$ are given by
\begin{align}
	p_a &= \cos^2(\beta' l' \Delta\gravP/2)\,,
	&
	p_b &= \sin^2(\beta' l' \Delta\gravP/2)\,,
\end{align}
see also Ref.~\cite{2021arXiv211115591M} for a similar analysis of entangled photon pairs (with finite photon bandwidths).

Compared to the setup discussed in the previous section, this interferometer has the advantage that both photons take the same path when traveling between the two unbalanced Mach--Zehnder interferometers, i.e., the two interferometers with unequal arm lengths labeled by $\hat a', \hat b'$ and $\hat a'', \hat b''$ in \Cref{fig:schematics TBI}. Thus, the vertical contributions of the photon trajectories automatically contribute equal phases to both parts of intermediate superposition states, while the layout sketched in \Cref{fig:schematics Mach--Zehnder} requires special precautions to ensure that the two vertical parts of the photon trajectories produce equal phase shifts.

\section{Discussion}

In this paper we have developed a general formalism for the Gupta--Bleuler quantization of the electromagnetic field in linear dielectrics, in terms of the gauge potential, valid also in (static) curved space-times.
The methods developed were applied to optical fibers placed horizontally in weak gravitational fields, yielding a consistent first-principles description of gravitational quantum optics experiments aiming at measuring gravitationally induced phase shifts on either single photons or entangled multi-photon states \cite{2012CQGra..29v4011R,2017NJPh...19c3028H,2021arXiv211115591M,2022PhRvA.106c1701M}.

The description of gravitational influences on light using the gauge potential given here could be useful for future comparison of similar effects in models with massive photons, such as the Proca equation.

Using the methods described here, it is also possible to describe optical fibers aligned arbitrarily in a uniform gravitational field, as well as allowing for fibers of arbitrary shape (though the equations become more intricate).
While the focus of this paper was on the perturbation of light by weak gravitational fields, the formalism developed here is not limited to the weak-field regime and could thus serve as a basis for studies of strong-field effects as well.

\section*{Acknowledgements}

My thanks go to Piotr Chruściel and Christopher Hilweg for helpful discussions.
T.M.\ is a recipient of a DOC Fellowship of the Austrian Academy of Sciences at the Faculty of Physics at the University of Vienna and is supported by the Vienna Doctoral School in Physics (VDSP), as well as the research platform TURIS.

\appendix

\section{Klein--Gordon Products of Wave-Packets}
\label{app:KG products wave-packets}

This section elaborates on the meaning of \Cref{eq:KG product modes general} by considering normalizable wave-packets constructed from the non-normalizable modes computed in \Cref{s:fibers in flat space}.

A positive-frequency wave-packet is defined to be an expression of the form
\begin{equation}
	\emA = \int_{\mathbf R} \dd \beta \sum_{I} \sum_{m \in \mathbf Z} \sum_{\kappa}
	\alpha_{I,\beta,m,\kappa} \emA_{I,\beta,m,\kappa}
\end{equation}
where $\emA_{I,\beta,m,\kappa}$ are the fiber modes computed in \Cref{s:fibers in flat space} (see \Cref{s:fiber KG product} for details on the notation), and where
\begin{equation}
	\int_{\mathbf R} \dd \beta \sum_{I} \sum_{m \in \mathbf Z} \sum_{\kappa}
	|\alpha_{I,\beta,m,\kappa}|^2 < \infty.
\end{equation}
Using the factorization \eqref{eq:fiber mode factorisation A P}, one finds the Klein--Gordon product \eqref{eq:fiber KG product} of two such positive-frequency wave-packets $A$ and $A'$ to be given by
\begin{equation}
	\begin{split}
		\kgbracket{A | A'}
		= & i \int_{\mathbf R} \hspace{-0.5em} \dd z
		\int_{[0,2\pi]} \hspace{-1.5em} \dd \theta
		\int_{\mathbf R_+} \hspace{-1.0em} \dd r \, r \!\!
		\int_{\mathbf R} \hspace{-0.5em} \dd \beta
		\int_{\mathbf R} \hspace{-0.5em} \dd \beta'
		\sum_{I,I'}
		\sum_{m,m'}
		\sum_{\kappa,\kappa'}\\
		& e^{i(\beta' - \beta) z}
		e^{i(m' - m) \theta}
		e^{i(\omega - \omega') t}\\
		& \times \bar \alpha_{I,\beta,m,\kappa}
		\alpha_{I',\beta',m',\kappa'}'\\
		& \times
		\{
			\bar a_{I,\beta,m,\kappa}(r) \cdot \pi_{I',\beta',m',\kappa'}(r)\\
			&\quad
			- 
			a_{I',\beta',m',\kappa'}(r) \cdot \bar\pi_{I,\beta,m,\kappa}(r)
		\}\,.
	\end{split}
\end{equation}
Using Plancherel’s theorem for the Fourier transform and Parseval’s theorem for the Fourier series, one obtains
\begin{equation}
	\begin{split}
		\kgbracket{A | A'}
		&= i (2 \pi)^2
		\int_{\mathbf R} \hspace{-0.5em} \dd \beta
		\sum_{m \in \mathbf Z}
		\int_{\mathbf R_+} \hspace{-1.0em} \dd r \, r
		\sum_{I,I'}
		\sum_{\kappa,\kappa'}\\
		&\quad e^{i(\omega - \omega') t}
		\bar \alpha_{I,\beta,m,\kappa}
		\alpha_{I',\beta,m,\kappa'}'\\
		&\quad\times
		\{
			\bar a_{I,\beta,m,\kappa}(r) \cdot \pi_{I',\beta,m,\kappa'}(r)\\
			&\qquad
			- a_{I',\beta,m,\kappa'}(r) \cdot \bar\pi_{I,\beta,m,\kappa}(r)
		\}\,.
	\end{split}
\end{equation}
One thus arrives at the formula
\begin{multline}
	\kgbracket{A | A'}
		= \int_{\mathbf R} \hspace{-0.5em} \dd \beta
		\int_{\mathbf R} \hspace{-0.5em} \dd \beta'
		\sum_{I,I'}
		\sum_{m,m'}
		\sum_{\kappa,\kappa'}\\
		\bar \alpha_{I,\beta,m,\kappa}
		\alpha_{I',\beta',m',\kappa'}'
		\kgbracket{A_{I,\beta,m,\kappa} | A_{I',\beta',m',\kappa'}}\,,	
\end{multline}
where
\begin{multline}
	\kgbracket{A_{I,\beta,m,\kappa} | A_{I',\beta',m',\kappa'}}
		= i (2 \pi)^2 \delta_{m, m'} \delta(\beta - \beta')\\
		\times e^{i(\omega - \omega') t} \int_0^\infty \hspace{-1.0em} \dd r \, r \{
			\t a{^*_\mu} \t \pi{^\prime^\mu}
			- \t a{^\prime_\mu} \t \pi{^*^\mu}
		\}\,.
\end{multline}
Here, the indices $(I,\beta,m,\kappa)$ of $a$ and $\pi$, as well as the indices $(I',\beta',m',\kappa')$ of $a'$ and $\pi'$ have been suppressed for brevity.

The extension to wave-packets which also have negative-frequency components is straightforward: for wave-packets of the form
\begin{multline}
	\emA = \int_{\mathbf R} \hspace{-0.5em} \dd \beta \sum_{I} \sum_{m \in \mathbf Z} \sum_{\kappa}\\
	\left\{
		\alpha^{(+)}_{I,\beta,m,\kappa} \emA_{I,\beta,m,\kappa}
		+ \alpha^{(-)}_{I,\beta,m,\kappa} \emA_{I,\beta,m,\kappa}^*
	\right\}\,,
\end{multline}
which contain both positive and negative frequency contributions, one obtains
\begin{equation}
	\begin{split}
		&\kgbracket{A | A'}
		= \int_{\mathbf R}\hspace{-0.5em} \dd \beta
		\int_{\mathbf R}\hspace{-0.5em} \dd \beta'
		\sum_{I,I'}
		\sum_{m,m'}
		\sum_{\kappa,\kappa'}\\
		\bigg\{
			&\bar \alpha^{(+)}_{I,\beta,m,\kappa}
			\alpha'^{(+)}_{I',\beta',m',\kappa'}
			\kgbracket{A_{I,\beta,m,\kappa} | A_{I',\beta',m',\kappa'}}\\
			&- \bar \alpha^{(-)}_{I,\beta,m,\kappa}
			\alpha'^{(-)}_{I',\beta',m',\kappa'}
			\kgbracket{A_{I,\beta,m,\kappa} | A_{I',\beta',m',\kappa'}}^*
		\bigg\}\,,
	\end{split}
\end{equation}
which is in accordance with the general relations
\begin{align}
	\kgbracket{A | A^*} &= 0\,,
	&
	\kgbracket{A^* | A'^*} &= - \kgbracket{A | A'}^*\,.
\end{align}

\section{Mode Orthogonality}
\label{app:orthogonality fiber modes}

In this section, it is shown that any two fiber-optic modes (as constructed in \Cref{s:optical fibers}) are orthogonal in the sense of the Klein--Gordon product (defined in \Cref{s:KG product}) whenever their azimuthal indices, propagation constants, or frequencies differ.

For a linear isotropic dielectric of constant permeability $\permeability$, which is inertial in flat space-time, the Klein--Gordon product \eqref{eq:KG product EM general} in the Feynman--’t~Hooft gauge $\xi = \permeability$, with the gauge parameter set to $\alpha = 1$, takes the form
\begin{multline}
	\kgbracket{A | A'}
		= \frac{i}{\permeability} \int \big\{
			n^2 \t\gopt{^a^b}(
				 \t*\emA{^*_a} \t\p{_0} \t*\emA{^\prime_b}
				- \t*\emA{^\prime_a} \t\p{_0} \t*\emA{^*_b}
			)\\
			+ n^2 \t\nabla{^i}(
				\t*\emA{^*_0} \t*\emA{^\prime_i}
			 - \t*\emA{^\prime_0} \t*\emA{^*_i}
		 )
		\big\}\dd V\,.
\end{multline}
For modes of the form given in \cref{eq:potential decomposition}, one obtains
\begin{multline}
	\label{eq:orthogonality modes product abstract}
	\kgbracket{A | A'}
		= (2 \pi)^2 \permeability^{-1} \delta_{m,m'} \delta(\beta - \beta')\\
		\times e^{i(\omega- \omega')t} \kgbracket{a | a'}_\text{red.}\,,
\end{multline}
where the reduced Klein--Gordon product splits into bulk and interface contributions as 
\begin{align}
	\label{eq:orthogonality product split}
	\kgbracket{a | a'}_\text{red.}
		&= \kgbracket{a | a'}_\text{blk}
		+ \kgbracket{a | a'}_\text{int}\,,
		\\
	\shortintertext{wherein}
	\label{eq:orthogonality product bulk}
	\kgbracket{a | a'}_\text{blk}
		&= (\omega + \omega') \int_0^\infty \hspace{-1em} \dd r\, r\, n^2 \t\gopt{^a^b} \t*a{^*_a} \t*a{^\prime_b}\,,
		\\
	\label{eq:orthogonality product interface}
	\kgbracket{a | a'}_\text{int}
		&= i \rho \jump{n^2(\t*a{^*_0} \t*a{^\prime_r}- \t*a{^\prime_0} \t*a{^*_r})}\,.
\end{align}
As the product given in \cref{eq:orthogonality modes product abstract} vanishes if the propagation constants $\beta$ and $\beta'$, or the azimuthal indices $m$ and $m'$ differ, all following calculations will be carried out for $\beta = \beta'$ and $m = m'$ only.
Setting $\mathfrak b_\nu$ to be the Bessel operator
\begin{equation}
	\bessel_\nu = \p_r^2 + r^{-1} \p_r - r^{-2} \nu^2\,,
\end{equation}
and defining
\begin{multline}
	\underline\Bessel
	\begin{pmatrix}
		a_t &
		a_\parallel &
		a_+ &
		a_-
	\end{pmatrix}\\
	=
	n^{-2}
	\begin{pmatrix}
		\mathfrak b_m a_t &
		\mathfrak b_m a_\parallel &
		\mathfrak b_{m+1} a_+ &
		\mathfrak b_{m-1} a_-
	\end{pmatrix}\,,
\end{multline}
the radial field equations \eqref{eq:fiber radial equations} can be written in the form
\begin{equation}
	\label{eq:orthogonality eigenvalue equation}
	[\underline\Bessel + \omega^2 - \beta^2/n^2]
	\begin{pmatrix}
		a_t &
		a_\parallel &
		a_+ &
		a_-
	\end{pmatrix}
	= 0\,.
\end{equation}
For the considered solutions $a$ and $a'$, set
\begin{equation}
	\label{eq:orthogonality delta definition}
	\varDelta = (\omega^2 - \omega'^2) \kgbracket{a | a'}_\text{red.}\,.
\end{equation}
Decomposing the reduced product into bulk and interface contributions according to \cref{eq:orthogonality product split}, and using \cref{eq:orthogonality eigenvalue equation} in the bulk term, one obtains
\begin{multline}
	\varDelta
		= \kgbracket{a | \underline\Bessel a'}_\text{blk}
		- \kgbracket{\underline\Bessel a | a'}_\text{blk}\\
		+ (\omega^2 - \omega'^2) \kgbracket{a | a'}_\text{int}\,.
\end{multline}
Integrating the bulk terms by parts and using the definition of $\kgbracket{a | a'}_\text{int}$ given in \cref{eq:orthogonality product interface}, one arrives at
\begin{equation}
	\label{eq:orthogonality delta expanded}
	\begin{split}
		\varDelta
		&= \rho (\omega + \omega') \jump{
			- n^2 \Wronskian(a_0^*, a_0')
			+ \Wronskian(a_\parallel^*, a_\parallel')
		}\\
		&\quad
		+ \rho (\omega + \omega') \jump{
			\Wronskian(a_+^*, a_+')
			+ \Wronskian(a_-^*, a_-')
		}\\
		&\quad
		+ i \rho (\omega^2 - \omega'^2) \jump{
			n^2(\t*a{^*_0} \t*a{^\prime_r}- \t*a{^\prime_0} \t*a{^*_r})
		}\,,
	\end{split}
\end{equation}
where $\Wronskian$ denotes the Wronskian
\begin{equation}
	\Wronskian(f,g) = f \p_r g - g \p_r f\,,
\end{equation}
and $\jump{f}$ denotes the jump of a function $f$ at the core-cladding interface:
\begin{equation}
	\jump{f} = \left( \lim_{r \nearrow \rho} f(r) \right) - \left( \lim_{r \searrow \rho} f(r) \right)\,.
\end{equation}

So far, no interface conditions on the fields $a$ and $a'$ were used.
Assuming the field to satisfy the matching conditions listed in \Cref{tab:matching gauge-fixed special}, one finds
\begin{subequations}
\begin{align}
	\jump{a_t'} &= 0\,,
	&
	\jump{n^2 \p_r a_t'} &= -i \omega' \jump{n^2} a_r'\,,
	\\
	\jump{a_r'} &= 0\,,
	&
	\jump{\p_r a_r'} &= - i \omega' \jump{n^2} a_0'\,,
	\\
	\jump{a_\theta'} &= 0\,,
	&
	\jump{\p_r a_\theta'} &= 0\,,
	\\
	\jump{a_z'} &= 0\,,
	&
	\jump{\p_r a_z'} &= 0\,,
\end{align}
and analogous equations for $a^*$:
\begin{align}
	\jump{a_t^*} &= 0\,,
	&
	\jump{n^2 \p_r a_t^*} &= +i \omega \jump{n^2} a_r^*\,,
	\\
	\jump{a_r^*} &= 0\,,
	&
	\jump{\p_r a_\theta^*} &= 0\,,
	\\
	\jump{a_\theta^*} &= 0\,,
	&
	\jump{\p_r a_r^*} &= + i \omega \jump{n^2} a_0^*\,,
	\\
	\jump{a_z^*} &= 0\,,
	&
	\jump{\p_r a_z^*} &= 0\,.
\end{align}
\end{subequations}
Since the $z$ components of the fields and their first radial derivatives are continuous, one has
\begin{equation}
	\label{eq:orthogonality jump W_z}
	\jump{\Wronskian(a_\parallel^*, a_\parallel')} = 0\,.
\end{equation}
Next, decomposing $\jump{\Wronskian(a_+^*, a_+') + \Wronskian(a_-^*, a_-')}$ into $r$ and $\theta$ components by means of \cref{eq:fiber complex frame}, using the continuity of the $\theta$ components and their first derivatives, as well as the continuity of the $r$ components, and also using the formulae for the jumps of the radial derivatives of the $r$ components, one finds
\begin{multline}
	\label{eq:orthogonality jump W_r}
	\jump{\Wronskian(a_+^*, a_+') + \Wronskian(a_-^*, a_-')}\\
	= \jump{\Wronskian(a_r^*, a_r')}
	= - i \jump{n^2} (\omega a_0^* a_r' + \omega' a_0' a_r^*)\big|_{\rho}\,.
\end{multline}
Finally, using the continuity of the $t$ components and the formula for the jump of its radial derivative, one obtains
\begin{equation}
	\label{eq:orthogonality jump W_t}
	\jump{n^2 \Wronskian(a_0^*, a_0')} = - i \jump{n^2} (\omega' a_0^* a_r' + \omega a_0' a_r^*)\big|_{\rho}\,.
\end{equation}
Inserting \cref{eq:orthogonality jump W_z,eq:orthogonality jump W_r,eq:orthogonality jump W_t} into \cref{eq:orthogonality delta expanded}, one finds $\varDelta = 0$, so that \cref{eq:orthogonality delta definition} yields
\begin{equation}
	(\omega^2 - \omega'^2) \kgbracket{a | a'}_\text{red.} = 0\,.
\end{equation}
This shows that modes of different frequencies are orthogonal in the sense of the Klein--Gordon product.
\Cref{eq:orthogonality modes product abstract} then entails that the Klein--Gordon product of two modes is always time-independent.

\section{Normalization Factors}
\label{app:fiber normalization}

This section contains explicit calculations of the normalization factors $N_{\beta,m,\kappa}$ and $N'_{\beta,m,\kappa}$ of the fiber modes given in \cref{eq:mode physical q,eq:mode gauge q,eq:mode ghost q}.

\subsection{Physical Modes}
\label{s:fiber normalization physical}
For the physical modes described in \Cref{s:fiber modes physical}, \cref{eq:fiber KG product} yields
\begin{multline}
	\kgbracket{\emA_{\phys \beta,m,\kappa} | \emA_{\phys \beta', m',\kappa'}}\\
		= (2\pi)^2 \delta(\beta - \beta') \delta_{m,m'} \delta_{\kappa,\kappa'} \frac{2 \rho^2 \beta^2 \omega}{|N_{\beta,m,\kappa}|^2} I_1\,,
\end{multline}
where
\begin{equation}
	\label{eq:KG norm physical integrals}
	\begin{split}
		I_1
		&= I_1(m,\beta,\omega,n_1,n_2,\rho)\\
		&= \frac{n_1^2}{2 U^2}
			\left(1 + \frac{\tilde m \omega/\beta}{\Jj + \Kk}\right)
			\left(n_1^2 + \frac{\tilde m \beta/\omega}{\Jj + \Kk}\right)
			\\&\quad\times
			\int_0^\rho \frac{J_{m+1}(U r/\rho)^2}{J_m(\rho)^2} r \, \dd r \\
		&+ \frac{n_1^2}{2 U^2}
			\left(1 - \frac{\tilde m \omega/\beta}{\Jj + \Kk}\right)
			\left(n_1^2 - \frac{\tilde m \beta/\omega}{\Jj + \Kk}\right)
			\\&\quad\times
			\int_0^\rho \frac{J_{m-1}(U r/\rho)^2}{J_m(\rho)^2} r \, \dd r \\
		&+ \frac{n_2^2}{2 W^2}
			\left(1 + \frac{\tilde m \omega/\beta}{\Jj + \Kk}\right)
			\left(n_2^2 + \frac{\tilde m \beta/\omega}{\Jj + \Kk}\right)
			\\&\quad\times
			\int_\rho^\infty \frac{K_{m+1}(W r/\rho)^2}{K_m(W)^2} r\, \dd r \\
		&+ \frac{n_2^2}{2 W^2}
			\left(1 - \frac{\tilde m \omega/\beta}{\Jj + \Kk}\right)
			\left(n_2^2 - \frac{\tilde m \beta/\omega}{\Jj + \Kk}\right)
			\\&\quad\times
			\int_\rho^\infty \frac{K_{m-1}(W r/\rho)^2}{K_m(W)^2} r\, \dd r\,.
	\end{split}
\end{equation}
These integrals can be calculated analytically using the formulae
\begin{align}
	\begin{split}
		\int_0^\rho \hspace{-0.25em} & \dd r\, r\, J_\nu(U r /\rho)^2\\
			&= + \half \rho^2 \left(
				J_\nu(U)^2 - J_{\nu+1}(U) J_{\nu-1}(U)
			\right)\,,
	\end{split}
	\\
	\begin{split}
		\int_\rho^\infty \hspace{-0.5em} & \dd r\, r\ K_\nu(W r /\rho)^2\\
			&= - \half \rho^2 \left(
				K_\nu(W)^2 - K_{\nu+1}(W) K_{\nu-1}(W)
			\right)\,.
	\end{split}
\end{align}
Numerically, $I_1$ is found to be positive (this was checked for all physical modes depicted in the mode diagram of \Cref{fig:mode diagrams}), so that the normalization of \cref{eq:fiber KG norm phys} is achieved by setting
\begin{equation}
	N_{\beta,m,\kappa}
		= 2 \pi \rho \beta \sqrt{2 \omega I_1}\,.
\end{equation}

\subsection{Gauge Modes and Ghost Modes}
\label{s:fiber normalization gauge ghost}

In \Cref{s:fiber modes gauge,s:fiber modes ghost}, it was shown that the gauge and ghost modes have vanishing Klein--Gordon norm. For their mutual product, \cref{eq:fiber KG product} yields
\begin{multline}
	\kgbracket{\emA_{\ghost \beta, m,\kappa} | \emA_{\gauge \beta', m',\kappa'}}\\
		= (2 \pi)^2 \delta(\beta - \beta') \delta_{m,m'} \delta_{\kappa,\kappa'} \frac{2 \rho^2 \beta^2 \omega}{|N'_{\beta, m,\kappa}|^2} I_2\,,
\end{multline}
where
\begin{equation}
	\begin{split}
		I_2
		&= \frac{2 n_1^2}{\rho^2} \int_0^\rho \frac{J_m(Ur/\rho)^2}{J_m(U)^2} \,r\, \dd r\\
		&\quad + \frac{2 n_2^2}{\rho^2} \int_\rho^\infty \frac{K_m(Wr/\rho)^2}{K_m(W)^2} \,r\, \dd r
		\\
		&= n_1^2 U^2 \Jj^2
		 + n_2^2 W^2 \Kk^2\\
		 & \quad
		 + (n_1^2 - n_2^2)
		 \left(
			 1 - \frac{m^2 \rho^2 \beta^2}{U^2 W^2}
		 \right)\,.
	\end{split}
\end{equation}
The first line shows that $I_2$ is positive, so that the normalization of \cref{eq:fiber KG prod gauge ghost} is attained by
\begin{equation}
	N'_{\beta,m,\kappa}
		= 2 \pi \rho \beta \sqrt{2 \omega I_2}\,.
\end{equation}

\section{Reduction of the Metric to Newtonian Form}
\label{app:metric reduction PPN to N}

In this section it is shown, for linear gravitational potentials, that all linearized parametrized-post-Newtonian (PPN) metrics of the form given in \cref{eq:metric PPN} can be reduced to the Newtonian form \eqref{eq:metric N} by a spatial coordinate transformation and rescaling of the gravitational potential by a constant.

Consider the PPN metric
\begin{equation}
	g
		= - (1 + 2 \aLPI \gravP) \dd t^2
		+ (1 - 2 \gamma \gravP) (\dd x^2 + \dd y^2 + \dd z^2)\,,
\end{equation}
where the gravitational potential $\gravP$ is an affine function of $z$:
\begin{equation}
	\gravP = \gravP_0 + \gravA z\,.
\end{equation}
To formulate the transformation to new coordinates $x'$, define the vector field
\begin{equation}
	\t\xi{^i}(x,y,z)
		= \begin{bmatrix}
			x z	&
			y z &
			\half (z^2 - x^2 - y^2)
		\end{bmatrix}\,,
\end{equation}
which satisfies
\begin{equation}
	\t\p{_i} \t\xi{_j} + \t\p{_j} \t\xi{_i}
		= 2 z \t\delta{_i_j}\,.
\end{equation}
Now, the coordinate transformation
\begin{equation}
	\t x{^i} = \t{{x'}}{^i} + \gamma \gravA\, \t\xi{^i}(x')
\end{equation}
brings the metric to the form
\begin{subequations}
\begin{align}
	\t*g{^\prime_0_0} &= - (1 + 2 \aLPI \gravP)\,,
	\\
	\t*g{^\prime_0_i} &= 0\,,
	\\
	\begin{split}
		\t*g{^\prime_i_j}
		&= (1 - 2 \gamma \gravP_0 + O(\gamma^2 \gravP_0 \gravP))\t\delta{_i_j}\\
		&\quad+ (1-2 \gamma \gravP) O(\gamma^2 \gravA^2 |x'|^2)\,,
	\end{split}
\end{align}
\end{subequations}
or, if error terms are neglected (which is admissible for sufficiently weak gravitational fields with $\gravP \ll 1$ and sufficiently small spatial regions $\gravA |x'| \ll 1$):
\begin{equation}
	g
		= - (1 + 2 \aLPI \gravP) \dd t^2
		+ (1 - 2 \gamma \gravP_0) (\dd x'^2 + \dd y'^2 + \dd z'^2)\,.
\end{equation}
Further, rescaling the coordinates according to
\begin{equation}
	\t{{x''}}{^i}
		= \sqrt{1 - 2 \gamma \gravP_0}\, \t{{x'}}{^i}\,,
\end{equation}
one obtains
\begin{equation}
	g
		= - (1 + 2 \aLPI \gravP) \dd t^2
		+ \dd x''^2 + \dd y''^2 + \dd z''^2\,.
\end{equation}
This is of the same form as the Newtonian metric
\begin{equation}
	g_N
		= - (1 + 2 \gravP) \dd t^2
		+ \dd x''^2 + \dd y''^2 + \dd z''^2\,,
\end{equation}
with $\gravP$ rescaled by the LPI parameter $\aLPI$.
This calculation shows that the PPN parameter $\gamma$ is irrelevant for experiments for which the approximation of linearized and uniform gravitational fields suffices.

\bibliography{bibliography}
\end{document}